\documentclass[pra,twocolumn,showpacs,superscriptaddress,amssymb,floatfix]{revtex4}
\usepackage[dvips]{graphicx}
\usepackage{bm}
\newcommand{\Journal}[4]{#1 {\bf #2}, #3 (#4)}
\newcommand{\PR}{Phys. Rev.}
\newcommand{\PRL}{Phys. Rev. Lett.}
\newcommand{\PRA}{Phys. Rev. A}
\newcommand{\JMP}{J. Math. Phys.}
\newcommand{\EPJD}{Eur. Phys. J. D}

\newcommand{\RMP}{Rev. Mod. Phys.}
\newcommand{\PLA}{Phys. Lett. A}
\newcommand{\NJP}{New Journal of Physics}
\newcommand{\Physica}{Physica}
\newcommand{\PPMSJ}{Proc. Phys. Math. Soc. Japan}
\newcommand{\AUW}{Acta Univ. Wratislaviensis}
\begin{document}
\title {Effective interactions, Fermi-Bose duality, and ground states of 
ultracold atomic vapors in tight de Broglie waveguides}
\author{M. D. Girardeau}
\email{girardeau@optics.arizona.edu}
\affiliation{Optical Sciences Center, University of Arizona, Tucson, AZ 85721}
\author{Hieu Nguyen}
\affiliation{Department of Physics and Astronomy, University of Southern 
California, Los Angeles, CA 90089-0484}
\author{M. Olshanii}
\email{olshanii@phys4adm.usc.edu}
\affiliation{Department of Physics and Astronomy, University of Southern 
California, Los Angeles, CA 90089-0484}

\date{\today}
\begin{abstract}
Derivation of effective zero-range one-dimensional (1D) interactions between
atoms in tight waveguides is reviewed, as is the Fermi-Bose mapping
method for determination of exact and strongly-correlated many-body
ground states of ultracold bosonic and fermionic atomic vapors in
such waveguides, including spin degrees of freedom. Odd-wave 1D interactions
derived from 3D p-wave scattering are included as well as the usual
even-wave interactions derived from 3D s-wave scattering, with emphasis
on the role of 3D Feshbach resonances for selectively enhancing s-wave
or p-wave scattering so as to reach 1D confinement-induced resonances
of the even and odd-wave interactions.
A duality between 1D fermions and bosons with zero-range interactions 
suggested by Cheon and Shigehara is shown to hold for the effective 
1D dynamics of a spinor Fermi gas with both even and odd-wave interactions
and that of a spinor Bose gas with even and odd-wave interactions,
with even(odd)-wave Bose coupling constants inversely related to 
odd(even)-wave Fermi coupling constants. Some recent applications of
Fermi-Bose mapping to determination of many-body ground states of Bose gases 
and of both magnetically
trapped, spin-aligned and optically trapped, spin-free Fermi gases are 
described, and a new generalized Fermi-Bose mapping is used to determine 
the phase diagram of ground-state total spin of the spinor Fermi gas 
as a function of its even and odd-wave coupling constants.
\end{abstract}
\maketitle

\section{\label{section:intro}Introduction}
The spin-statistics theorem, according to which identical particles
with integer spin are bosons whereas those with half-integer spin are
fermions, breaks down if the particles are confined to one or two
dimensions. Realization of this fact had its origin over forty years ago
when it was shown \cite{Gir60,Gir65} that the many-body problem of hard-sphere
bosons in one dimension can be mapped exactly onto that of an ideal Fermi gas,
so that many properties of such Bose systems are Fermi-like. It is now
known that this ``Fermi-Bose duality'' is a very general property of
identical particles in 1D, not restricted to the hard-sphere model,
and relating strongly interacting bosons to weakly-interacting fermions
and vice versa. In recent years this esoteric subject has become highly
relevant through experiments on ultracold atomic vapors in atom waveguides 
\cite{Key,Muller,Dek00,Thy,Bongs,Denschlag,Greiner,Arlt,GorVogLea01}. 
An understanding of their properties is 
important for atom interferometry \cite{Ber97,Dum02} and integrated atom 
optics \cite{Sch98,Dek00,ReiThy03}, which are potentially important for
development of ultrasensitive detectors of accelerations and gravitational
anomalies. 

When an ultracold atomic vapor is placed into an atom waveguide with
sufficiently tight transverse confinement, its two-body scattering
properties are strongly modified and short-range correlations are
greatly enhanced. This occurs in a regime of low temperatures
and densities where both the chemical potential $\mu$ and the thermal
energy $k_{B}T$ are less than the transverse oscillator level spacing 
$\hbar\omega_{\perp}$, so transverse oscillator modes are frozen and the 
dynamics is described by a one-dimensional (1D) Hamiltonian with zero-range 
interactions \cite{Ols98,PetShlWal00}. 
This 1D regime has already been reached 
experimentally \cite{Greiner,MorStoKohEss03,Tol03}, as has a regime 
with $\mu<\hbar\omega_{\perp}$ but $k_{B}T>\hbar\omega_{\perp}$ 
\cite{Bongs,GorVogLea01}.
We assume herein that both $\mu<\hbar\omega_{\perp}$ and 
$k_{B}T<\hbar\omega_{\perp}$ as in \cite{Greiner,MorStoKohEss03,Tol03}.
Nevertheless, \emph{virtually} excited transverse 
modes renormalize the effective 1D coupling constant via a confinement-induced
resonance, as first shown for bosons \cite{Ols98,BerMooOls03} and recently 
for spin-polarized fermionic vapors \cite{GraBlu03}.
At the very low densities of ultracold atomic vapors, the 3D interatomic
interactions are usually adequately described by the s-wave scattering length
and a corresponding 3D zero-range pseudopotential, the input for 
the original derivation \cite{Ols98} of an effective 1D interaction between 
waveguide-confined spinless bosonic atoms. A more detailed and comprehensive
theory has been developed recently \cite{MooBerOls04}. In the simplest case of 
spinless bosons the resultant effective 1D interaction is of the form 
$g_{1D}^B\delta(z)$ where $g_{1D}^B$ is an explicit and
nontrivial function \cite{Ols98,BerMooOls03,MooBerOls04} 
of the 3D s-wave scattering length,
$z=z_{1}-z_{2}$, and $z_1$ and $z_2$ are 1D coordinates of the interacting
atoms measured along the longitudinal axis of the waveguide. This derivation
\cite{Ols98,MooBerOls04} will be described in Sec. \ref{subsec:scattering},
and in Sec. \ref{subsec:pseudopotentials} the derivation of effective 1D
pseudopotentials will be presented, for application
not only to the spinless Bose gas but also to the case of spinor Fermi and
Bose gases, for which the definition of pseudopotentials is much more
delicate due to wave function discontinuities induced in the zero-range
limit by 1D odd-wave interactions derived from 3D p-wave scattering.
In the spatially uniform case (no longitudinal trapping potential) the 
zero-temperature properties of an $N$-atom spinless Bose gas are determined 
by a dimensionless coupling
constant $\gamma_{B}=mg_{1D}^B/n\hbar^2$ \cite{LieLin63,Ols98,PetShlWal00} 
where $m$ is the atomic mass and $n=N/L$ is the 1D number density, $L$ being 
the length of the periodic box. The exact $N$-boson ground state was found 
in the spatially uniform case by the
Bethe Ansatz method in a famous paper of Lieb and Liniger (LL) 
\cite{LieLin63}, and spawned later development of a powerful and more general 
approach \cite{KorBogIze93}. An exact solution in the presence of longitudinal 
trapping is not known, but is well approximated by a local equilibrium
approach \cite{DunLorOls01}. 

Since the density is in the denominator, at sufficiently 
\emph{low} densities one enters the regime of strong interactions and 
strong short-range correlations where effective-field approaches fail, exactly
the opposite of the situation in 3D where the density is in the numerator
of the dimensionless coupling constant. For $\gamma_{B}\gg 1$ the scattering 
reduces to specular reflection (negligible transmission coefficient)
and the Hamiltonian reduces to that of impenetrable point bosons, whose exact 
$N$-particle ground state
was found in 1960 by an exact mapping to the \emph{ideal} Bose gas 
\cite{Gir60}, leading to ``fermionization'' of many properties of the
Bose gas and related to breakdown of the symmetrization postulate and
the spin-statistics theorem \cite{Gir65}. This Fermi-Bose mapping method
and its application to determination of exact N-atom ground states
will be described in Sec. \ref{subsec:FBa}. 
In Sec. \ref{subsec:FBb} a very powerful generalization of this mapping
to arbitrary coupling strength \cite{CheShi98} will be described and its
application to determination of the ground state of the magnetically trapped,
spin-aligned Fermi gas will be reviewed, and in Sec. \ref{subsec:FBc} 
application of two further generalizations of the mapping to determination of 
the ground state of the optically trapped spinor Fermi gas will be 
described. 
\section{\label{sec:Atomic}Atomic scattering and 1D interactions in tight 
waveguides}
\subsection{\label{subsec:scattering}Scattering theory in tight waveguides}
\subsubsection{Formulation of the scattering problem}
\label{subsec:formulation}
We begin from the Hamiltonian for two atoms under transverse harmonic 
confinement and
subject to an arbitrary interaction potential
\begin{eqnarray}
\label{H2atoms}
\hat{H}_2&=&-\frac{\hbar^2}{2m_1}\nabla^2_1-\frac{\hbar^2}{2m_2}\nabla^2_2
    +\frac{1}{2}m_1\omega_\perp^2{\bf r}^2_{1\perp}
    +\frac{1}{2}m_2\omega_\perp^2{\bf r}^2_{2\perp}\nonumber\\
    &+&V({\bf r}_1-{\bf r}_2)
\end{eqnarray}
where $m_1$ and $m_2$ are the atomic masses, $\omega_\perp$ is the 
transverse trap
frequency, and $\nabla^2_i$ and ${\bf r}_{i\perp}$ are the Laplacian and 
radial coordinate
of the $i^{th}$ atom, respectively. This Hamiltonian is separable in 
relative and
center-of-mass coordinates ${\bf R}=(m_1{\bf r}_1+m_2{\bf r}_2)/M$, ${\bf 
r}={\bf r}_1-{\bf
r}_2$, $M=m_1+m_2$ being the total mass, yielding 
$\hat{H}_2=\hat{H}_{rel}+\hat{H}_{COM}$,
where
\begin{equation}
\label{Hrel}
    \hat{H}_{rel}=-\frac{\hbar^2}{2\mu}\nabla^2_{\bf 
r}+\frac{1}{2}\mu\omega_\perp^2{\bf r}^2_\perp
    +V({\bf r}),
\end{equation}
and
\begin{equation}
\label{HCOM}
    \hat{H}_{COM}=-\frac{\hbar^2}{2M}\nabla^2_{\bf 
R}+\frac{1}{2}M\omega_\perp^2{\bf R}^2_\perp,
\end{equation}
where $\mu=m_1m_2/(m_1+m_2)$ is the reduced mass, and ${\bf r}_\perp$ and 
${\bf R}_\perp$
are the relative and center-of-mass radial coordinates, respectively. The 
center-of-mass
Hamiltonian is that of a simple harmonic oscillator whose solution is 
known, hence we focus
only on the relative motion of the two particles. This reduces the problem 
to a single
particle of mass $\mu$, subject to transverse harmonic confinement, which 
is scattered by
an external potential $V({\bf r})$. The central equation which must be 
solved is therefore
Schr\"odinger's equation for the state of relative motion of the two atoms
\begin{equation}
\label{HpsiE}
    \left[E-\hat{H}-\hat{V}\right]|\psi(E)\rangle =0
\end{equation}
where $\hat{H}=\hat{H}_{rel}$ and $\hat{V}$ is the interatomic potential. 
Determining the
eigenstates of this Hamiltonian, in particular within the s-wave scattering 
approximation
for $V({\bf r})$, is the central goal of this section.
\subsubsection{Definitions and theorems}
\label{subsec:definitions}
In this subsection we briefly review the T-matrix formulation of scattering 
theory, which
provides a convenient framework for approaching the present problem. Let us 
first introduce
the retarded Green's function for a system with the Hamiltonian $\hat{H}$ 
and energy $E$
\begin{equation}
\label{GEH}
    \hat{G}_{\hat{H}}(E)
    = \lim_{\epsilon\to 0^+}(E + i\epsilon - \hat{H})^{-1}.
\end{equation}
We then define the T-matrix at energy $E$ of the scatter $\hat{V}$ in the 
presence of the
background Hamiltonian $\hat{H}$ in 
the usual manner as
\begin{eqnarray}
\label{TEofVG}
\hat{T}_{\hat{H},\hat{V}}(E)&=&\left[1-\hat{V}\hat{G}_{\hat{H}}(E)\right]^{-1}
    \hat{V}\nonumber\\
    &=&\sum_{n=0}^\infty \left[\hat{V}\hat{G}_{\hat{H}}(E)\right]^n\hat{V},
\end{eqnarray}
the summation form being valid provided that there are no difficulties with 
convergence.

Two relations on which we will rely heavily are the Lippman-Schwinger 
relation
\begin{equation}
\label{LippSchwing}
    \hat{G}_{\hat{H}+\hat{V}}(E)=\hat{G}_{\hat{H}}(E)
    +\hat{G}_{\hat{H}}(E)\hat{T}_{\hat{H},\hat{V}}(E)\hat{G}_{\hat{H}}(E),
\end{equation}
which relates the full Green's function of the system $\hat{H}+\hat{V}$ to 
the unperturbed
Green's function $\hat{G}_{\hat{H}}(E)$ and the T-matrix, and the Lupu-Sax 
formula \cite{Lupu-Sax}
\begin{eqnarray}
\label{LupuSax}
&&\hat{T}_{\hat{H},\hat{V}}(E)
\\
&&\quad
 =
    \left[1-\hat{T}_{\hat{H}',\hat{V}}(E)
    \left[\hat{G}_{\hat{H}}(E)-\hat{G}_{\hat{H}'}(E)\right]
    \right]^{-1}\hat{T}_{\hat{H}',\hat{V}}(E),
\nonumber
\end{eqnarray}
which relates the T-matrix of the scatter $\hat{V}$ in the background 
Hamiltonian $\hat{H}$
to the T-matrix for the same scatter but in a different background 
Hamiltonian $\hat{H}'$.
\subsubsection{Scattering theory}
\label{subsec:scattering_theory}
In the continuous part of the spectrum of the total Hamiltonian 
$\hat{H}+\hat{V}$
its eigenstates can be expressed as a sum of
an incident and a scattered wave according to
\begin{equation}
\label{psispis0}
    |\psi(E)\rangle=|\psi_0(E)\rangle+|\psi_s(E)\rangle,
\end{equation}
where $|\psi_0(E)\rangle$, the ``incident'' state vector, satisfies
\begin{equation}
\label{Gpsi0}
   \hat{G}^{-1}_{\hat{H}}(E)|\psi_0(E)\rangle=0,
\end{equation}
we can then express the Schr\'odinger equation for the total system as
\begin{equation}
\label{GVpsi}
    \left[\hat{G}^{-1}_{\hat{H}}(E)-\hat{V}\right]
    \left(|\psi_0(E)\rangle+|\psi_s(E)\rangle\right)=0.
\end{equation}
This equation is readily solved for the scattered wave in terms of the 
unperturbed Green's
function and the T-matrix, yielding
\begin{equation}
\label{TEHV}
|\psi_s(E)\rangle=\hat{G}_{\hat{H}}(E)\hat{T}_{\hat{H},\hat{V}}(E)|\psi_0(E
)\rangle,
\end{equation}
which will serve as the basis for our treatment of the present scattering 
problem.
\subsubsection{s-wave scattering regime: the reference T-matrix approach}
\label{swavepseudopot}
At first glance it may seem that finding the T-matrix is no easier than a 
direct solving of
the Schr\"odinger equation (\ref{HpsiE}). We will demonstrate, however, 
that the T-matrix
formulation allows for a self-consistent description of the low-energy part 
of the spectrum
that uses the {\it free-space} low-energy scattering properties of the 
interaction potential
as the {\it only} input. In addition the low-energy (s-wave) limit is 
isolated to a single
well-defined approximation without requiring the ad-hoc introduction of 
regularization via a
pseudo-potential. In this section we first outline this self-consistent 
low-energy
treatment. We then solve for the T-matrix using the standard Huang-Fermi 
pseudo-potential,
showing that the pseudo-potential reproduces the exact result in this 
situation.

Let the unperturbed Hamiltonian $\hat{H}$ be a Hamiltonian for a single 
nonrelativistic
particle in presence of a trapping potential $U$:
\begin{eqnarray}
\label{Hgeneral}
    \langle{\bf r}|\hat{H}|\psi\rangle
    =\left[ -\frac{\hbar^2 \nabla^2_{{\bf r}}}{2\mu} + U({\bf 
r})\right]\langle{\bf r}|\psi\rangle \quad ,
\end{eqnarray}
Assume also that the particle is `perturbed' by a scatterer given by
\begin{eqnarray}
\label{Vgeneral}
    \langle{\bf r}|\hat{V}|\psi\rangle
    =V({\bf r})\langle{\bf r}|\psi\rangle \quad
\end{eqnarray}
localized around ${\bf r} = {\bf 0}$. In what follows we will derive a {\it 
low-energy
approximation} for the T-matrix of the scatterer $V$ in presence of 
$\hat{H}$. It is
important to note that, by definition, the T-matrix acts only on 
eigenstates of the
unperturbed Hamiltonian, which we can safely assume to be regular 
everywhere. (This is of
course a constraint on the properties of the unperturbed Hamiltonian.) In 
this case the
zero-range s-wave scattering limit does not require any regularization of 
the T-matrix. By
making use of the Lupu-Sax formula (\ref{LupuSax}), we first derive the 
correct form of the
T-matrix in the low-energy s-wave regime without the introduction of a 
regularized
pseudo-potential. In the following section, however, we will see that the 
results we obtain
are in agreement with the standard Huang-Fermi pseudopotential approach to 
s-wave
scattering.

We begin our derivation by first specifying a ``reference'' background 
Hamiltonian $\hat{H}'$
as
\begin{eqnarray}
\label{H'}
    \langle{\bf r}|\hat{H}'|\psi\rangle
    =\left[-\frac{\hbar^2 \Delta_{{\bf r}}}{2\mu}+E\right]\langle{\bf 
r}|\psi\rangle.
\end{eqnarray}
This Hamiltonian is that of a free particle, but with an explicit energy 
dependence included
so that the eigenstates have zero wavelength at all energies. We note that 
this reference
Hamiltonian agrees with the free-space Hamiltonian in the zero-energy 
limit. While this
Hamiltonian may seem strange, it is a valid reference Hamiltonian which 
turns out to be
useful because the resulting T-matrix is energy independent for any 
scattering potential.
The Green's function for this Hamiltonian is given by
\begin{eqnarray}
\label{G'}
     \langle {\bf r} | \hat{G}_{\hat{H}'}(E) | {\bf r}' \rangle =
     -\frac{\mu}{2\pi \hbar^2}
    \frac{1}{|{\bf r}-{\bf r}^{\prime}|},
\end{eqnarray}
as can be verified by direct substitution into $[E-\hat{H}']\,
\hat{G}_{\hat{H}'}(E)=\hat{I}$. In turn the $T$-matrix of the interaction 
potential $V$ in
presence of $\hat{H}'$ is independent of energy and can therefore be 
expressed as
\begin{equation} \label{T'}
     \langle {\bf r}|\hat{T}_{\hat{H}',\hat{V}}(E)|{\bf r}'\rangle
    =g D({\bf r},{\bf r}'),
\end{equation}
where the kernel $D$ is defined as normalized to unity,
\begin{equation}
\label{Dnorm}
    \int d{\bf r} d{\bf r}' \, D({\bf r},{\bf r}') = 1.
\end{equation}
The normalization coefficient $g$ is then related, through the 
zero-energy scattering amplitude, to the three-dimensional 
scattering length $a_{s}$
according to
\begin{equation}
    g=\frac{2\pi\hbar^2 a_{s}}{\mu}.
\label{g}
\end{equation}

Imagine that the kernel $D({\bf r},{\bf r}')$ is well localized within some 
radius $R$. In
perturbative expansions at low energies this kernel only participates in 
convolutions with
slow (as compare to $R$) functions, in which case it can be approximated by 
a
$\delta$-function,
\begin{equation}
\label{Ddelta}
    D({\bf r},{\bf r}') \approx \delta({\bf r}) \delta({\bf r}').
\end{equation}
This straightforward approximation is the key to the s-wave scattering 
approximation. 
This effectively replaces 
the exact
reference T-matrix by its long-wavelength limit, so that the reference 
T-matrix assumes the
form
\begin{equation}
\label{T'delta}
    \langle {\bf r}|\hat{T}_{\hat{H}',\hat{V}}(E)|{\bf r}'\rangle
    \stackrel{k,k^{\prime} \ll 1/R}{\approx} g \delta({\bf r})\delta({\bf r}'),
\end{equation}
which is equivalent to
\begin{equation}
\label{TH'E}
    \hat{T}_{\hat{H}',\hat{V}}(E)=g|0\rangle\langle 0|,
\end{equation}
where $|0\rangle$ is the position eigenstate corresponding to the location 
of the scatterer.
In expression (\ref{T'delta}) $k$ and $k'$ refer to the wavevectors of any 
matrices which
multiply the T-matrix from the left and right, respectively.

If we now substitute the above expression for the reference T-matrix into 
the Lupu-Sax
formula (\ref{LupuSax}) for the T-matrix under the background Hamiltonian 
$\hat{H}$ we
arrive at
\begin{eqnarray}
\label{expansion_1}
&&    \hat{T}_{\hat{H},\hat{V}}(E)
\\
&&\quad    =\sum_{n=0}^\infty \left[g|0\rangle\langle 
0|\hat{G}_{\hat{H}'}(E)\right]^ng|0\rangle\langle 0|
\nonumber
\\
&&\quad    =
    \left[1-g\langle 0|\hat{G}_{\hat{H}}(E)|0\rangle+g\langle 
0|\hat{G}_{\hat{H}'}(E)|0\rangle\right]^{-1}
    g|0\rangle\langle 0|
    .
\nonumber
\end{eqnarray}
Making use of Eq. (\ref{G'}), we introduce the function $\chi({\cal E})$, 
defined as
\begin{eqnarray}
\label{chi}
    \chi(E)=\lim_{{\bf r}\to 0}\left[\langle{\bf 
r}|\hat{G}_{\hat{H}}(E)|0\rangle
    +\frac{\mu}{2\pi\hbar^2|{\bf r}|}\right],
\end{eqnarray}
from which we obtain the following simple expression for the T-matrix of 
the scatterer
$\hat{V}$ in presence of the trap:
\begin{equation}
\label{Tzerorange}
    \langle {\bf r} | \hat{T}_{\hat{H},\hat{V}}(E) | \psi \rangle
    \stackrel{E \ll \hbar^2/\mu R^2}{\approx}
    \frac{g\delta({\bf r})}{1-g\chi(E)}
    \langle {\bf r}
    | \psi \rangle.
\end{equation}

From comparing the equations the free-space and bound Green's functions obey
one can show that
the
singularity in bound Green's function is the same as that in the free-space 
Green's
function. Hence, $\chi(E)$ is the value of the regular part of the bound 
Green's function at
the origin.

For the case of transverse harmonic confinement function $\chi(E)$ 
has been explicitly computed in \cite{MooBerOls04}. It reads
\begin{eqnarray}
\chi(E)=-\frac{\mu}{2\pi a_\perp}\zeta(1/2,-(\frac{E}{2\hbar\omega_\perp}-\frac{1}{2})),
\label{gchiEfinal}
\end{eqnarray}
where $\zeta(s,\alpha)$ is 
the generalized Riemann zeta function described in the mathematical literature 
\cite{Hurwitz}:
\begin{eqnarray}
\label{zeta}
&&    \zeta(s,\alpha)=\lim_{N\to\infty}
\left[
\left(
    \sum_{n=0}^N \frac{1}{(n+\alpha)^{s}}
\right)
    - \frac{1}{1-s} \, \frac{1}{(N+\alpha)^{s-1}}
\right]
\nonumber
\\
&&\mbox{Re}(s)>0, \quad -2\pi < \arg(n+\alpha) \le 0.
\end{eqnarray}
Note that no established convention for choosing the branch 
of the irrational power functions exist: the choice above 
is just the most suitable for the needs of this paper.
\subsubsection{\label{E1D}Effective one-dimensional interaction potential 
for waveguide-confined spinless bosons}
By a direct substitution to the equation for the Green's function 
of the relative motion of two particles in a waveguide 
\begin{eqnarray} 
(E-\hat{H}+i\epsilon)\langle{\bf r}|\hat{G}_{\hat{H}}(E)|0\rangle = \delta({\bf r})
\end{eqnarray}
it is easy to show that the Green's function can be decomposed 
to a sum over the transverse modes in the following way:
\begin{eqnarray} 
&&\langle{\bf r}|\hat{G}_{\hat{H}}(E)|0\rangle
\\
&&\quad = \sum_{n=0}^{\infty} \langle z| \hat{G}_{1D}(E-\hbar\omega_{\perp}(2n+1)) |0\rangle
\phi_{n}({\bm \rho})\phi_{n}^{*}(0),
\nonumber
\label{green_decomposition}
\end{eqnarray}   
where ${\bm \rho} = x {\bf e}_x + y {\bf e}_y$, 
$\phi_{n}({\bm \rho})$ are the zero-angular-momentum eigenstates 
of the transverse oscillator, and $\hat{G}_{1D}(E_{1D})$ is the 
Green's function of a free one-dimensional particle:
\begin{eqnarray}  
&&(E_{1D} + \frac{\hbar^2}{2\mu}\frac{\partial^2}{\partial z^2} + i\epsilon)
\langle z| \hat{G}_{1D}(E-\hbar\omega_{\perp}(2n+1)) |0\rangle 
\nonumber
\\
&&\quad
=
\delta(z) 
\end{eqnarray}    
assume now that our energy belongs to the single mode window: 
\begin{eqnarray}   
\hbar\omega_{\perp} \le E < 3\hbar\omega_{\perp}.
\end{eqnarray}     
In this case all the $n\ne 0$ terms in the expansion (\ref{green_decomposition})
exponentially decay at large $z$. Accordingly the solution of the scattering problem 
(\ref{psispis0}), (\ref{TEHV}) now resembles a solution of a 
{\it one-dimensional} scattering problem: 
\begin{eqnarray}
&&\psi({\bf r})
\stackrel{|z|\to \infty}{=}
\left\{
\psi_{0,\,1D}(z) + 
\right.
\\
&&\quad
\left.
\langle z| \hat{G}_{1D}(E_{1D}(E)) |0\rangle |\phi_{0}(0)|^2 \tau_{3D}(E) 
\psi_{0,\,1D}(0)
\right\}
\phi_{0}({\bm \rho}),
\nonumber
\label{tmp10}
\end{eqnarray}
where $\tau_{3D}(E)$ is the strength of the three-dimensional T-matrix, i.e.,
\begin{eqnarray}
\tau_{3D}(E) = \frac{g\delta({\bf r})}{1-g\chi(E)},
\end{eqnarray}   
and $E_{1D}(E) = E - \hbar\omega_{\perp}$. Comparing the expression 
(\ref{tmp10}) and the general form of a scattering solution 
(\ref{psispis0}), (\ref{TEHV})
it is natural to interpret the expression in the braces in the l.h.s.\ of (\ref{tmp10})
as a solution of a {\it one-dimensional} scattering problem, subject to a scattering 
potential whose (one-dimensional) T-matrix reads
\begin{eqnarray}
\langle z| \hat{T}_{1D}(E_{1D}) |0\rangle =
\tau_{1D}(E_{1D}) \delta(z)
\label{T1D}
\end{eqnarray}
with 
\begin{eqnarray} 
\tau_{1D}(E_{1D}) = |\phi_{0}(0)|^2 \tau_{3D}(\hbar\omega_{\perp} + E_{1D}). 
\label{tau1Dexact}
\end{eqnarray}   
For low energies $E_{1D} \ll \hbar\omega_{\perp}$ the strength of the 
one-dimensional T-matrix is approximately 
\begin{eqnarray} 
\tau_{1D}(E_{1D}) \stackrel{E_{1D} \ll \hbar\omega_{\perp}}{\approx} 
-\frac{\hbar^2}{\mu} \, 
\frac{1}
{
-\frac{a_{\perp}^2}{2a_{s}} \left[1 + \frac{a_{s}}{a_{\perp}}\zeta(1/2) \right] -\frac{i}{k}
}
\label{tau1Dapprox}
\end{eqnarray}   
The relative error of this approximation scales as ${\cal O}(k^3)$.

One can now attempt to introduce an effective one-dimensional scatterer whose T-matrix is close 
or equal to the one given above (\ref{T1D}), (\ref{tau1Dapprox}). 
A straightforward calculation shows that the T-matrix of a one-dimensional 
delta-potential 
\begin{eqnarray}  
\hat{V}_{1D} = g_{1D}^{B} \delta(z)
\label{delta-potential}
\end{eqnarray}
has a form 
\begin{eqnarray}
\tau_{1D,\delta}(E_{1D}) = 
-\frac{\hbar^2}{\mu} \,
\frac{1}
{
a_{1D}^B -\frac{i}{k}
},
\label{tau1Ddelta}
\end{eqnarray}   
where the one-dimensional scattering length
$a_{1D}^B$ is related to the potential strength by
\begin{eqnarray}
g_{1D}^{B} = -\frac{\hbar^2}{\mu a_{1D}^B}.
\end{eqnarray}
Comparison of the T-matrices 
(\ref{tau1Dapprox}) and (\ref{tau1Ddelta}) leads to a conclusion that 
the waveguide scattering T-matrix (\ref{tau1Dapprox}) can be exactly reproduced 
by a delta-potential (\ref{delta-potential}) of a scattering length
\begin{eqnarray}\label{Bose-a1D}
a_{1D}^B = - \frac{a_{\perp}^2}{2a_{s}} \left[1 + \frac{a_{s}}{a_{\perp}}\zeta(1/2) \right]
,
\end{eqnarray}
where $\zeta(x)$ is the Riemann zeta function and $\zeta(1/2)=-1.4603\ldots$.

The above expression reproduces the result of \cite{Ols98}
where the effective one-dimensional potential has been obtained via 
a straightforward solution of the scattering problem. Notice 
that at $a_{s} = |\zeta(1/2)|^{-1} a_{\perp}$ the coupling constant diverges, 
signifying the so-called ``confinement induced resonance'' (CIR). 
The significance of this 
resonance has been confirmed in {\it ab initio} two-body numerical 
calculations 
with finite-range realistic interatomic potentials \cite{BerMooOls03}
and in many-body Monte-Carlo simulations \cite{AstBluGioGra03}.

Further extensions of the zero-range model for the effective 
one-dimensional scatterer can be envisioned. For example, if one chooses 
to to reproduce the scattering properties (more precisely the 
denominator of the one-dimensional T-matrix (\ref{tau1Dexact})) 
with a relative error of ${\cal O}(k^5)$, the one-dimensional delta-potential can 
be replaced by a rectangular potential of a finite width $2l$ and hight/depth $v_{0}$. 
In the case of 
repulsive interaction the model potential is a rectangular barrier, and 
for attractive interactions it is a rectangular well. In the limit of 
$a_{s} \ll a_{\perp}$ their half-widths $l$  are given by
\begin{eqnarray*}
\begin{array}{lll}
l = \frac{\sqrt{\zeta(3/2)}}{2} \, \sqrt{a_{s} a_{\perp}} 
  = 0.8081 \, \sqrt{a_{s} a_{\perp}} 
&\mbox{for}& g_{1D}^B > 0
\\
l = \frac{(\zeta(3/2)/2)^{2/3}}{18} \, a_{s} 
  = .0664 \, |a_{s}|
&\mbox{for}& g_{1D}^B < 0
\end{array},
\end{eqnarray*}
and in both cases the strength $v_{0}$ of the potential is 
\begin{eqnarray} 
v_{0} = |g_{1D}^B/2l| 
.
\end{eqnarray}   
%
\subsubsection{\label{subsubsec:p-wave}Remarks on p-wave scattering}
Attempts to carry through a program analogous to the above for 
the case of p-wave scattering between polarized fermions meet  
numerous (hopefully technical) obstacles: Most of the limiting 
procedures become mutually nonuniformly convergent and no clear way 
to  identify a correct order 
is visible. In any case the closest candidate 
for the p-wave analog of the free-space three-dimensional T-matrix 
is the pseudopotential introduced in \cite{Roth} 
\begin{equation} 
     \langle \chi |\hat{T}_{\hat{H}',\hat{V}}(E)| \psi \rangle
    = \frac{27\pi\hbar^2 V_{p}}{\mu}\bm{\nabla}\chi^{*}(0) \cdot \bm{\nabla}\psi(0)
\end{equation}
that can be shown to reproduce correctly the low-energy behavior 
of the p-wave scattering amplitude. Here $V_{p}$ is the p-wave scattering 
volume,
that defines the low-energy behavior of the p-wave scattering phase via 
$V_{p} = -\lim_{k\to 0} \tan \delta_{p}(k)/k^3$ \cite{SunEsrGre03}. 

An elegant way around these difficulties has been fond recently 
by Granger and Blume \cite{GraBlu03}, who used a K-matrix technique  
that does not 
explicitly involve any zero-range objects. The analysis 
of polarized Fermi gases presented below is heavily based on the Granger and 
Blume findings. 
\subsection{\label{subsec:pseudopotentials}1D pseudopotentials}
\subsubsection{\label{subsubsec:spinless-bosons}Spinless bosons}
This is the simplest case. 
The 1D scattering length $a_{1D}^B$ is defined in terms of the ratio of 
derivative $\psi_{B}^{'}(z)$
and value $\psi_{B}(z)$ of the relative wave function 
just outside the range $z_0$ of the interaction:
\begin{equation}\label{eq1}
\psi_{B}^{'}(z_0)
=-\psi_{B}^{'}(-z_0)=-(a_{1D}^{B}-z_{0})^{-1}\psi_{B}(\pm z_0)\quad ,
\end{equation}
which is equivalent in the zero-range limit $z_{0}\to 0+$ to the familiar LL 
\emph{contact condition} \cite{LieLin63,KorBogIze93} 
\begin{equation}\label{Bose-contact}
\psi_{B}^{'}(0+)=-\psi_{B}^{'}(0-)
=\frac{\mu g_{1D}^{B}}{\hbar^2}\psi_{B}(0\pm)
\end{equation}
for the delta function interaction $g_{1D}^{B}\delta(z)$ provided that
the scattering length and coupling constant are related by 
$g_{1D}^{B}=-\hbar^{2}/\mu a_{1D}^B$ where $\mu$ is the effective mass 
$m/2$. It follows from the expression for $a_{1D}^{B}$ derived in 
\cite{Ols98,MooBerOls04} and Eq.(\ref{Bose-a1D}) herein that 
\begin{equation}\label{CIR-Bose}
g_{1D}^{B}=2a_{s}\hbar\omega_{\perp}\left[1
-\frac{a_s}{a_{\perp}}|\zeta(1/2)|\right]^{-1}
\end{equation}
implying the existence of a confinement-induced resonance CIR 
\cite{Ols98,BerMooOls03} of the coupling constant as $a_s$ is tuned via
a 3D Feshbach resonance \cite{Rob01} past the resonance point
$a_{s}/a_{\perp}=|\zeta(1/2)|^{-1}=0.6848\ldots$. Hence the whole range
of 1D coupling constants from $-\infty$ to $+\infty$ is experimentally
achievable by tuning $a_s$ over a narrow range in the neighborhood of the 1D
resonance. It was shown recently \cite{BerMooOls03} that this 
is a 1D Feshbach resonance between ground and excited transverse vibrational
manifolds. It was shown in \cite{BerMooOls03} and in Sec. \ref{E1D} herein 
that at low longitudinal energies $ka_{\perp}\ll 1$ the 1D scattering 
amplitude generated by the 
interaction $g_{1D}^{B}\delta(z)$ reproduces the exact 3D scattering amplitude
in the waveguide to within a relative error ${\cal O}(k^3)$.
\subsubsection{\label{subsubsec:spin-aligned}Spin-aligned fermions}
Consider next a magnetically trapped, spin-aligned atomic vapor of
spin-$\frac{1}{2}$ fermionic atoms in a tight waveguide. The $N$-fermion spin 
wave function is magnetically frozen in the configuration 
$\uparrow_{1}\cdots\uparrow_{N}$, 
so the space-spin wave function must be \emph{spatially} antisymmetric,
s-wave scattering is forbidden, and the leading interaction effects at 
low energies are determined by the 3D p-wave scattering amplitude. 
Such p-wave interactions are usually negligible at the low densities of
ultracold atomic vapors, but they can be greatly enhanced by p-wave
Feshbach resonances \cite{RegTicBohJin03}.
Granger and Blume derived an effective one-dimensional K-matrix for 
the corresponding two-fermion problem \cite{GraBlu03} in a tight waveguide. 
In the low-energy \cite{Note1} domain the K-matrix 
can be reproduced, with a relative error ${\cal O}(k^{3})$, 
by the contact condition \cite{GraBlu03,GirOls03}
\begin{equation}\label{Fermi-contact}
\psi_{F}(0+)=-\psi_{F}(0-)= -a_{1D}^{F}\psi_{F}^{'}(0\pm)
\end{equation}
where 
\begin{eqnarray}\label{Fermi-renorm}
a_{1D}^{F}&=&\frac{6V_{p}}{a_{\perp}^2}[1+12(V_{p}/a_{\perp}^3)
|\zeta(-1/2,1)|]^{-1} 
\end{eqnarray}
is the odd-wave one-dimensional scattering length, 
$V_{p}=a_{p}^{3}=-\lim_{k\to 0}\tan\delta_{p}(k)/k^3$ is the 
p-wave ``scattering volume'' \cite{SunEsrGre03}, $a_{p}$ is the p-wave
scattering length, and
$\zeta(-1/2,1)=-\zeta(3/2)/4\pi=-0.2079\ldots$ is 
the Hurwitz zeta function evaluated at 
$(-1/2,1)$ \cite{WhiWat52}. 
The expression (\ref{Fermi-renorm}) has a resonance
at a \emph{negative} critical value  
$V_{p}^{crit}/a_{\perp}^{3}=-0.4009\cdots$.
Note that $a_{1D}^{F}\to -\infty$ as $V_p$ approaches this critical
value, implying that the exterior wave function (i.e., outside the interaction
region $|z|<z_{0}\to 0$) satisfies the \emph{free-particle} Schr\"{o}dinger
equation. This is the opposite of the bosonic case, where it follows
from Eq. (\ref{Bose-a1D}) that $a_{1D}^{B}\to 0$ at resonance.

In accordance with (\ref{Fermi-contact}),  
the low-energy fermionic wavefunctions, Eq. (20)
of \cite{GraBlu03}, are discontinuous at 
contact, but left and right limits of their derivatives coincide. 
$V_p$ is tunable via a 3D Feshbach resonance \cite{RegTicBohJin03},
allowing experimental realization of all values of $a_{1D}^F$ from
$-\infty$ to $+\infty$. It will be shown that in this fermionic case
the effective 1D coupling constant is 
$g_{1D}^{F}=-\hbar^{2}a_{1D}^{F}/\mu$, which can be compared and contrasted
with the previously defined bosonic 1D coupling constant 
$g_{1D}^{B}=-\hbar^{2}/\mu a_{1D}^B$. The dimensionless fermionic coupling
constant is $\gamma_{F}=mg_{1D}^{F}n/\hbar^2$. Note that the density $n$
is in the numerator, whereas it is in the denominator of the bosonic analog
$\gamma_{B}=mg_{1D}^{B}/n\hbar^2$. For $\gamma_{F}\gg 1$ one has a 
``fermionic TG gas'' \cite{GirOls03}, 
a fermionic analog of the impenetrable Bose gas,
called the ``Tonks-Girardeau'' (TG) gas in recent 
literature \cite{DunLorOls01,OlsDun02,OhbSan02,OlsDun03,GanShl03,AstGio02, 
BogMalBulTim03,LieSeiYng03,ReiThy03,KheGanDruShl03,MorStoKohEss03,DruDeuKhe03,
BerBorIzrSme04,AstBluGioGra03,AstBluGioPit03,Ger04}. As previously noted, 
$a_{1D}^{F}\to -\infty$ in the fermionic TG limit, implying an 
interaction-free exterior wave function. It will be shown in 
Sec. \ref{subsec:FBb} that in this limit the Fermi gas maps to an 
\emph{ideal Bose} gas, providing a physical explanation of the 
interaction-free nature of the exterior wave function. This can be compared 
and contrasted with the bosonic TG gas, which maps to an \emph{ideal Fermi}
gas.

\begin{figure}
\includegraphics[width=1.0\columnwidth,angle=0]{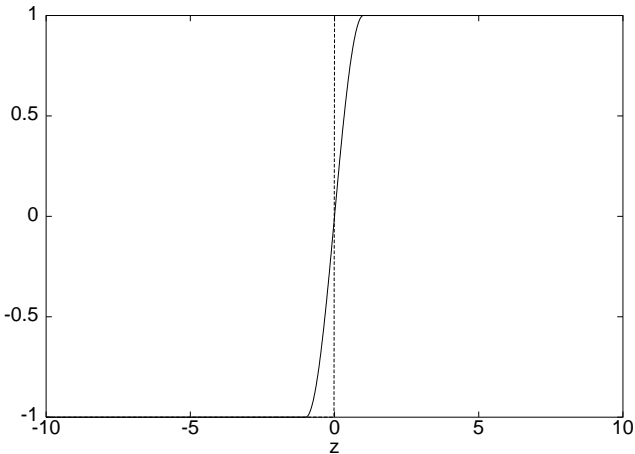}
\caption{$N=2$ untrapped fermionic TG gas ground state (dashed line) 
compared with zero-energy scattering solution for a square well 
with range $z_0$ and depth $V_0$ corresponding to the boundary between no 
bound state and one bound state, a zero energy resonance (solid line), 
as function of relative coordinate z. Units are such that $z_0=1$.} 
\label{Fig:one}
\vspace{-0.5cm}
\end{figure}

Although a discontinuity in the derivative 
is a well-known consequence of the zero-range delta function pseudopotential
and plays a crucial role in the solution of the Lieb-Liniger model
\cite{LieLin63}, discontinuities of $\psi$ itself have received 
little attention, although they have been discussed previously by
Cheon and Shigehara \cite{CheShi98} and are implicit in the recent work
of Granger and Blume \cite{GraBlu03}. For a fermionic wave function $\psi_F$ 
the discontinuity $2\psi_{F}(0+)$ is a trivial consequence of antisymmetry 
together with the fact that a nonzero odd-wave
scattering length cannot be obtained in the limit $z_{0}\to 0$ unless 
$\psi_{F}(0\pm)\ne 0.$ These discontinuities are rounded off when $z_{0}>0$,
since the interior wave function interpolates smoothly between the values
at $z=-z_0$ and $z=z_0$. This is illustrated in Fig. \ref{Fig:one} for the
special case of the fermionic TG gas, the limit $\gamma_{F}\gg 1$ 
\cite{GirOls03}. The
potential is chosen to be a square \emph{well} because we will find later
that stability of the ground state against collapse requires that the
corresponding effective zero-range interaction ($z_{0}\to 0+$)
be \emph{negative} definite, which can be shown to be the case when 
$g_{1D}^{F}>0$ and hence both $V_{p}<0$ and $a_{1D}^{F}<0$.  
The energy is taken as zero so the exterior solution is 
$\text{sgn}(z)=\pm 1$; an interior
solution fitting smoothly onto this is $\sin(\kappa z)$ with
$\kappa=\sqrt{2\mu V_{0}/\hbar^2}=\pi/2z_0$, the critical
value where the last bound state passes into the continuum, 
a zero-energy resonance. A fermionic contact condition with a finite
scattering length can be obtained in the limit $z_{0}\to 0$ if $\kappa$
scales with the width $z_{0}$ as 
$\kappa=(\pi/2z_{0})[1+(2/\pi)^2 (z_{0}/a_{1D}^{F})]$.

Following the bosonic case, where 
the $\delta$-interaction can be introduced naturally to cancel 
the $\delta$-functions resulting from double differentiation of functions with 
discontinuous derivatives, in the case of fermions whose wave function is 
discontinuous it is tempting to introduce $\delta'$ interactions. However, 
$\delta'$-functions and second derivatives are ill-defined if used in a 
convolution with discontinuous functions. This difficulty is resolved by 
realizing
that (as will be shown), the $\delta'$ is associated with the \emph{interior}
wave function ($-z_{0}<z<z_0$) whereas the contact condition 
(\ref{Fermi-contact}) refers to the $z_{0}\to 0+$ limit of the wave function
at $z=\pm z_{0}\ne 0$, just \emph{outside} the range of the interaction
potential. Take the Hamiltonian to be 
$\hat{H}_{1D}^{F}=-(\hbar^{2}/2\mu)\partial_{z}^{2}+\hat{v}_{1D}^{F}$
where $\hat{v}_{1D}^{F}$ is a pseudopotential operator to be determined.
$\partial_{z}^{2}$ is nonsingular for $z\ne 0$,  
but at the origin there are singular contributions. The first 
derivative is
$\partial_{z}\psi_{F}(z)=\psi_{F}^{'}(z\ne 0)
+[\psi_{F}(0+)-\psi_{F}(0-)]\delta(z)$. The second derivative then has
an additional singular contribution from differentiation of the delta 
function:
\begin{equation}\label{Fermi-KE}
\partial_{z}^{2}\psi_{F}(z)=\psi_{F}^{''}(z\ne 0)
+[\psi_{F}(0+)-\psi_{F}(0-)]\delta^{'}(z)\quad .
\end{equation}
In the second term on the right, $\delta^{'}(z)$ is associated with the
interior wave function, as we will eventually verify explicitly from the
$z\to 0+$ limit of a finite-range interaction, whereas its prefactor arises
from the contact condition just outside the range of the interaction.
Define a linear regularizing operator $\hat{\partial}_{\pm}$ by
\begin{equation}\label{Fermi-regularize}
\hat{\partial}_{\pm}\psi_{F}(z)=(1/2)[\psi_{F}^{'}(0+)+\psi_{F}^{'}(0-)]\quad .
\end{equation}
Then the desired pseudopotential is 
$\hat{v}_{1D}^{F}=g_{1D}^{F}\delta^{'}(z)\hat{\partial}_{\pm}$. It satisfies
a convenient projection property $\hat{v}_{1D}^{F}\psi_{e}(z)=0$ where
$\psi_e$ is any even wave function. Although $\psi_F$ is always purely odd
(antisymmetric) for the spin-aligned Fermi gas, this projection property
is useful for the spinor (spin-free) Fermi gas to be discussed in
Sec. \ref{subsubsec:spin-free}, where the spatial dependence has, in general,
both even and odd contributions. This is the reason for the general form
$(1/2)[\psi_{F}^{'}(0+)+\psi_{F}^{'}(0-)]$ of the factor occurring in
Eq. (\ref{Fermi-regularize}), even though it could be simplified to
$\psi_{F}^{'}(0+)$ or $\psi_{F}^{'}(0-)$ when acting on $\psi_F$, for
which $\psi_{F}^{'}(0+)=\psi_{F}^{'}(0-)$ as a consequence of antisymmetry.
It is easy to show that terms in $\delta^{'}(z)$ cancel from
$\hat{H}_{1D}^{F}\psi_F$ as a consequence of the contact condition
(\ref{Fermi-contact}) provided that $g_{1D}^{F}=-\hbar^{2}a_{1D}^{F}/\mu$.

The physical significance is clarified by starting from the same square well
already discussed in connection with Fig. \ref{Fig:one}, i.e., $v(z)=-V_0$ 
when $-z_{0}<z<z_0$ and zero when $|z|>z_0$. The antisymmetric
solution $\psi_{F}$ of the zero-energy scattering equation
$[(-\hbar^{2}/2\mu)\partial_{z}^{2}+v(z)]\psi_{F}(z)=0$ inside the well 
is $\sin(\kappa z)$ with $\kappa=\sqrt{2\mu V_{0}/\hbar^2}$. The 
scattering length $a_{1D}^{F}$ is defined by  
$\psi_{F}(z_0)=-\psi_{F}(-z_0)=-(a_{1D}^{F}-z_{0})\psi_{F}^{'}(\pm z_0)$
which is satisfied in the limit $z_{0}\to 0+$ if $V_0$ scales with $z_0$ as 
$\kappa=(\pi/2z_{0})[1+(2/\pi)^2 (z_{0}/a_{1D}^{F})]$. In that limit the
boundary conditions reduce to 
Eq. (\ref{Fermi-contact}). Inside the well the kinetic and potential
energy terms are $-(\hbar^{2}/2\mu)\partial_{z}^{2}\psi_{F}(z)
=-(\hbar^{2}\kappa^{2}/2\mu)\sin(\kappa z)$ and 
$v(z)\psi_{F}(z)=-V_{0}\sin(\kappa z)$. For $|z|<z_0$,
$\cos(\kappa z)$ is proportional to a representation of $\delta(z)$
as $z_{0}\to 0$, since 
$\int_{-z_{0}}^{z_{0}}\cos(\kappa z)f(z)dz
\to f(0)\int_{-z_{0}}^{z_{0}}\cos(\kappa z)dz
=f(0)2\kappa^{-1}\sin(\kappa z_{0})\to 2z_{0}f(0)$.  
Then its derivative $-(\kappa/2z_{0})\sin(\kappa z)$
is a representation of $\delta^{'}(z)$. Noting that $\kappa z_{0}\to\pi/2$
as $z_{0}\to 0$ we have 
$-(\hbar^{2}/2\mu)\partial_{z}^{2}\psi_{F}(z)
=-(\hbar^{2}\kappa^{2}/2\mu)\sin(\kappa z)\to 
(\pi\hbar^{2}/2\mu)\delta^{'}(z)$ which agrees with the kinetic energy term
$-(\hbar^{2}/2\mu)[\psi_{F}(0+)-\psi_{F}(0-)]\delta^{'}(z)$ since 
$\psi_{F}(0+)$ and $\psi_{F}(0-)$ are to be interpreted as $\psi_{F}(z_{0})$ 
and $\psi_{F}(-z_{0})$ as $z_{0}\to 0+$. Next consider the
potential energy term inside the well as $z_{0}\to 0+$:
$-V_{0}\sin(\kappa z)\to -V_{0}(-2z_{0}/\kappa)\delta^{'}(z)
\to(\pi\hbar^{2}/2\mu)\delta^{'}(z)$. Comparing this with 
$\hat{v}_{1D}^{F}\psi_{F}(z)$, using the
expression for $g_{1D}^{F}$, noting that $\psi_{F}^{'}(0\pm)$ are to be 
interpreted as $\psi_{F}^{'}(\pm z_{0})$ for $z_{0}\to 0$,
one finds that the two expressions for the potential energy term agree
in that limit. It is clear from this derivation that the $\delta^{'}(z)$
in the effective 1D Hamiltonian can be interpreted as the ``ghost of the
vanished interior wave function'', which plays a crucial role in the odd-wave
interaction even in the limit $z_{0}\to 0$. 
\subsubsection{\label{subsubsec:spin-free}Spin-free fermions}
In this section we assume that the spinor Fermi gas is optically trapped,
so the spins are unconstrained. This case is more complicated than
the spin-aligned Fermi gas even in the absence of explicit
spin-spin interactions or external spin-dependent potentials, because
the requirement of antisymmetry under combined space-spin exchanges
$(z_{i},\sigma_{i})\leftrightarrow (z_{j},\sigma_{j})$ induces implicit
space-spin coupling leading to nontrivial spin dependence of the wave 
functions. Here each spin z-component argument $\sigma_i$ takes on the
values $\uparrow$ or $\downarrow$, or equivalently $\pm\frac{1}{2}$. Consider
3D two-body scattering in the waveguide. There are both s-wave scattering 
states, which are space symmetric and spin antisymmetric with spin 
eigenfunctions of singlet
form $\frac{1}{\sqrt{2}}(\uparrow\downarrow-\downarrow\uparrow)$, 
and p-wave scattering states, which are space antisymmetric 
and spin symmetric with spin eigenfunctions
of triplet form $\uparrow\uparrow$ or $\downarrow\downarrow$ or
$\frac{1}{\sqrt{2}}(\uparrow\downarrow+\downarrow\uparrow)$. 
Assume that the Hamiltonian does not depend on spin. Then the spin dependence 
of fermionic wave functions $\psi_F$ need not be
indicated explicitly and they can be written as the sum of  spatially even
and odd parts $\psi_e$ and $\psi_o$. The odd part decomposes further
into three components going with the three triplet spin eigenfunctions,
but one need not complicate the notation at this point since the Hamiltonian 
acts in the same way on 
each of them. The effective 1D interactions are determined by 1D
scattering lengths $a_{1D}^e$ for spatially even waves 
$\psi_{e}(z)=\psi_{e}(-z)$ related to 3D s-wave scattering and spatially
odd waves $\psi_{o}(z)=-\psi_{o}(-z)$ related to 3D p-wave scattering. 
The contact condition for 1D even-wave scattering is the same as the 
previously-given one (\ref{Bose-contact}) with $\psi_B$ and $a_{1D}^B$
replaced by $\psi_e$ and $a_{1D}^e$, and the one for 1D odd-wave scattering
is $\psi_{o}(z_0)=-\psi_{o}(-z_0)=-(a_{1D}^{o}-z_{0})\psi_{o}^{'}(\pm z_0)$,
the $z_{0}>0$ version of Eq. (\ref{Fermi-contact}), with $\psi_F$ and 
$a_{1D}^F$ replaced by $\psi_o$ and $a_{1D}^o$. In the zero-range limit
$z_{0}\to 0+$ these can be combined into \cite{GirOls03,GirOls04}
\begin{eqnarray}\label{General-contact}
\psi'(0+)-\psi'(0-)= -(a_{1D}^{e})^{-1}[\psi(0+) + \psi(0-)]\nonumber
\\
\psi(0+)-\psi(0-)= -a_{1D}^{o}[\psi'(0+) + \psi'(0-)]
\end{eqnarray}
where $\psi(z)=\psi_{e}(z)+\psi_{o}(z)$. In this section $\psi(z)$ is
a fermionic function $\psi_F$, but we use a more general notation 
because the same equations apply to a spinor Bose gas. 
$a_{1D}^e$ and $a_{1D}^{o}$
are related to the 3D s-wave scattering length $a_s$ and the 3D p-wave
scattering volume $V_p$ by Eqs. (\ref{Bose-a1D}) and (\ref{Fermi-renorm}), 
with $a_{1D}^B$ replaced by $a_{1D}^e$ and $a_{1D}^{F}$ replaced by 
$a_{1D}^{o}$. Take the Hamiltonian to be 
\begin{equation}\label{H_1D}
\hat{H}_{1D}=-(\hbar^{2}/2\mu)\partial_{z}^{2}+\hat{v}_{1D}^{e}
+\hat{v}_{1D}^{o}\quad .
\end{equation}
Here $\hat{v}_{1D}^{o}$ differs from $\hat{v}_{1D}^{F}$ of 
Sec. \ref{subsubsec:spin-aligned} only by the obvious substitutions, i.e.,
$\hat{v}_{1D}^{o}=g_{1D}^{o}\delta^{'}(z)\hat{\partial}_{\pm}$ with
$\hat{\partial}_{\pm}$ defined by Eq. (\ref{Fermi-regularize})
with $\psi_F$ replaced by $\psi_o$. The even-wave pseudopotential is 
more complicated than the simple delta-function interaction of
Sec. \ref{subsubsec:spinless-bosons}, because the delta function is ambiguous 
at the point $z=0$, where $\psi$ is discontinuous due to the discontinuity 
in its odd-wave component $\psi_o$. In order to determine the correct
form, start from the first derivative 
$\partial_{z}\psi(z)=\psi^{'}(z\ne 0)
+[\psi(0+)-\psi(0-)]\delta(z)$ as before. Now the second derivative 
has \emph{two} singular contributions in addition to the nonsingular term
$\psi^{''}(z\ne 0)$, one because in general 
$\psi^{'}(0+)\ne\psi^{'}(0-)$ and the other from the derivative of 
the delta function:
\begin{eqnarray}\label{KE}
\partial_{z}^{2}\psi(z)&=&\psi^{''}(z\ne 0)
+[\psi^{'}(0+)-\psi^{'}(0-)]\delta(z)\nonumber\\
&+&[\psi(0+)-\psi(0-)]\delta^{'}(z)\ .
\end{eqnarray}
With proper choice of $g_{1D}^{o}$ the odd-wave pseudopotential 
$\hat{v}_{1D}^{o}$ cancels the $\delta^{'}(z)$ term from the kinetic
energy, and we define the even-wave pseudopotential $\hat{v}_{1D}^{e}$
so as to cancel the $\delta(z)$ term: 
$\hat{v}_{1D}^{e}=g_{1D}^{e}\hat{\delta}_{\pm}$ where the linear operator
$\hat{\delta}_{\pm}$ is defined by 
\begin{equation}\label{even-interaction}
\hat{\delta}_{\pm}\psi(z)=(1/2)[\psi(0+)+\psi(0-)]\delta(z)\quad .
\end{equation}
These pseudopotentials satisfy convenient projection
properties $\hat{v}_{1D}^{e}\psi_{o}=\hat{v}_{1D}^{o}\psi_{e}=0$ on the even 
and odd parts of $\psi$, and their matrix elements are 
$\langle\chi|v_{1D}^{e}|\psi\rangle=\frac{1}{2}\chi^{*}(0)[\psi(0+)+\psi(0-)]$
and $\langle\chi|v_{1D}^{o}|\psi\rangle
=-\frac{1}{2}[\chi^{'}(0)]^{*}[\psi^{'}(0+)+\psi^{'}(0-)]$. They connect only 
even to even and odd to odd wave functions
if we stipulate that $\chi(0)=0$ [the average of $\chi(0+)$ and $\chi(0-)$]
if $\chi$ is odd and $\chi^{'}(0)=0$ [the average of
$\chi^{'}(0+)$ and $\chi^{'}(0-)$] if $\chi$ is even. In fact, the wave
function and its derivative at $z=0$ refer to the \emph{internal} wave
function as modified by the potential, whereas $z=0+$ and
$z=0-$ refer to the wave function \emph{just outside} the range of the
potential, and the above values at $z=0$ follow from the way the internal
wave function interpolates between the contact conditions on the
\emph{exterior} wave function. 
Using (\ref{General-contact}) one finds that 
terms in $\delta(z)$ and $\delta^{'}(z)$ cancel from
$\hat{H}_{1D}$ if the even and odd-wave coupling constants are related
to the scattering lengths by $g_{1D}^{e} = -\hbar^{2}/\mu a_{1D}^e$ and 
$g_{1D}^{o} = -\hbar^{2}a_{1D}^{o}/\mu$. 
\section{Fermi-Bose mapping methods and N-atom ground states}
In this section we will review the theory of Fermi-Bose mappings relating 
the exact $N$-particle energy eigenstates of systems of fermions and bosons
in 1D with effective zero-range interactions, and application of these
mappings to determination of the $N$-atom ground states. This will be 
done first for the original
mapping for impenetrable bosons ($\gamma\gg 1$) \cite{Gir60,Gir65}, then for a 
very powerful generalization to arbitrary values of $\gamma$ due to 
Cheon and Shigehara\cite{CheShi98}, 
and finishing with a further generalization to the case of spinor
Fermi and Bose gases, important for applications to optically trapped
fermionic atoms whose spins are unconstrained.
\subsection{\label{subsec:FBa}Impenetrable bosons (TG limit)}
It was already pointed out in the famous paper of Lieb and Liniger on the 
1D Bose gas with delta-function interactions \cite{LieLin63} and in
Secs. \ref{section:intro} and \ref{subsubsec:spin-aligned} 
that the 1D gas of impenetrable point bosons is the limit $\gamma_{B}\gg 1$
of the LL gas, the ``TG limit''. Tonks gave the first treatment of the
statistical mechanics of a 1D hard-sphere gas \cite{Ton36}, which
was restricted to the classical high-temperature regime 
and provided no information about the extreme quantum limit characteristic
of ultracold atomic vapors. The formula for the
exact quantum-mechanical ground-state energy of the 1D hard-sphere Bose gas
appeared in a paper of Bijl where it is quoted without derivation   
\cite{Bij37}, and a derivation was published by 
Nagamiya \cite{Nag40}. Then in 1960 one of us \cite{Gir60} and
Stachowiak \cite{Sta60} 
independently rederived the ground-state energy.
The Fermi-Bose mapping method was first introduced in \cite{Gir60},
although Nagamiya had previously noted \cite{Nag40} that in the 
``fundamental sector'' 
$z_{1}\le z_{2}\le\cdots\le z_{N}$ the ground state wave function of a 
spatially uniform, 1D hard-core Bose gas can be written as
an ideal Fermi gas determinant, continuation into other permutation sectors
being effected by imposing overall Bose symmetry under
all permutations $z_{i}\leftrightarrow z_j$ in spite of the fermionic
\emph{anti}symmetry under permutations of \emph{orbitals} (\emph{not} 
coordinates) in the fundamental sector. The mapping theorem is much more
general, also holding in the presence of external potentials and/or
finite two-particle or many-particle interactions in addition to the
hard core interaction \cite{Gir60}. It also applies to the
1D time-dependent many-body Schr\"{o}dinger equation and has been used
to treat some time-dependent interference properties of the 1D hard core 
Bose gas \cite{soliton,CCit,GirDasWri02,DasLapWri02,DasGirWri02}. 

We now briefly review the mapping theorem. The $N$-boson Hamiltonian is
assumed to have the structure 
$\hat{H}_{1D}=-(\hbar^2/2m)\sum_{j=1}^{N}\partial_{z_{j}}^{2}
+V(z_{1},\cdots,z_{N})$ where the real, symmetric function $V$ contains
all external potentials (e.g., a longitudinal trap potential) as well as
any finite interaction potentials \emph{not including} the hard-sphere
repulsion, which is instead treated as a constraint on allowed wave
functions $\psi_{B}(z_{1},\cdots,z_{N})$:
\begin{equation}\label{eq2}
\psi_{B}=0\quad\text{if}\quad |z_{j}-z_{k}|<a\quad,\quad 1\le j<k\le N \quad .
\end{equation}
Let
$\psi_{F}(z_{1},\cdots,z_{N})$ be a fermionic solution of 
$\hat{H}_{1D}\psi=E\psi$
which is antisymmetric under
all pair exchanges $z_{j}\leftrightarrow z_{k}$, hence all
permutations. One can consider $\psi_F$ to be either the wave function of
a fictitious system of ``spinless fermions'', or else that of a system of
real, spin-aligned fermions. Define a ``unit antisymmetric function" 
\cite{Gir60}
\begin{equation}\label{eq3}
A(z_{1},\cdots,z_{N})=\prod_{1\le j<k\le N}\text{sgn}(z_{k}-z_{j})  ,
\end{equation}
where $\text{sgn}(z)$ is the algebraic sign of the coordinate difference
$z=z_{k}-z_{j}$, i.e., it is +1(-1) if $z>0$($z<0$). For given
antisymmetric $\psi_F$,
define a bosonic wave function $\psi_B$ by
\begin{equation}\label{eq4}
\psi_{B}(z_{1},\cdots,z_{N})=A(z_{1},\cdots,z_{N})\psi_{F}(z_{1},\cdots,
z_{N})
\end{equation}
which defines the Fermi-Bose mapping. $\psi_B$ satisfies
the hard core constraint (\ref{eq2}) if $\psi_F$ does, is totally
symmetric (bosonic) under permutations, obeys the same
boundary conditions as $\psi_F$, and $\hat{H}_{1D}\psi_{B}=E\psi_{B}$ 
follows from
$\hat{H}_{1D}\psi_{F}=E\psi_{F}$ \cite{Gir60,Gir65}. In the case of periodic
boundary conditions (no trap potential, spatially uniform system) one must add
the proviso that the
boundary conditions are only preserved under the mapping if $N$ is odd,
but the case of even $N$ is accomodated by imposing periodic
boundary conditions on $\psi_F$ but {\em anti}periodic boundary conditions
on $\psi_B$. 

The mapping theorem leads to explicit expressions for all
many-body energy eigenstates and eigenvalues under the assumption that
the only two-particle interaction is a zero-range hard core
repulsion, represented by the $a\rightarrow 0$ limit of the
hard-core constraint, the ``TG gas''. Such solutions were obtained in Sec. 3 
of \cite{Gir60} 
for periodic boundary conditions and no external potential. At the low
densities of ultracold atomic vapors it is usually sufficient 
to consider this case, although it has been shown by
Astrakharchik {\it et al.} \cite{AstBluGioGra03} 
that for a longitudinally trapped LL gas
with attractive interaction $g_{1D}<0$ there is a regime within which
the equation of state is well approximated by taking $a>0$. 
The exact ground state is also known in this case 
\cite{Nag40,Gir60}. Here we limit ourselves to the usual 
strongly repulsive TG limit $\gamma_{B}\gg 1$. Since wave functions of 
``spinless'' or spin-aligned fermions are 
antisymmetric under coordinate exchanges,
their wave functions vanish automatically whenever any
$z_{j}=z_{k}$, the constraint has no effect, and the corresponding
fermionic ground state is the ground state of the {\em ideal} gas
of fermions, a Slater determinant of the lowest $N$
single-particle plane-wave orbitals. The exact many body ground
state was found \cite{Gir60} to have energy 
$E_{0}=(N-N^{-1})(\pi\hbar n)^{2}/6m$ where $n=N/L$ is the linear number 
density, and the wave function was found to be a pair product of 
Bijl-Jastrow form
\begin{equation}
\psi_{B0}=\text{const.}\prod_{i>j}|\sin[\pi L^{-1}(z_{i}-z_{j})]| ,
\end{equation}
where $L$ is the perimeter of the annular trap. In spite of the
very long range of the individual pair correlation factors
$|\sin[\pi L^{-1}(z_{i}-z_{j})]|$, the pair distribution function
$D(z_{ij})$, the joint probability density
that if one particle is found at $z_i$ a second will be found at
$z_j$, was found to be of short range
$D(z_{ij})=1- [\sin(\pi n z_{ij})/\pi n z_{ij}]^2$.
Clearly,
$D(0)=0$ which reflects the hard core nature of the two-particle
interaction. By examination of the excited states the system was
found to support propagation of sound with speed
$c=\pi\hbar n/m$ \cite{Gir60}, and it was shown that this agrees with the
thermodynamic formula in terms of the compressibility of the ground state.
``Fermionization'' holds only for those
properties expressible in terms of the configurational probability density
$|\psi_{B0}(z_{1},z_{2},\cdots,z_{N})|^{2}$. The momentum distribution depends
on the single-particle correlation function 
$g_{1}(z)=\langle\hat{\psi}^{\dagger}(z)\hat{\psi}(0)\rangle$ 
(reduced single-particle density matrix), which is very different from
that of the ideal Fermi gas and very difficult to evaluate. Its eigenfunctions
are plane waves $e^{ikz}$ because of translational invariance of the
system, and the corresponding eigenvalues define the momentum distribution
function $N(k)$, the discrete Fourier transform of $g_{1}(z)$, the allowed
values of $k$ being $k_{j}=j2\pi/L$ with $j=0,\pm 1,\pm 2,\cdots$. 
In a classic {\it tour de force} Lenard found
\cite{Len64,Len66} $N(0)$ to be of order $\sqrt{N}$, large but much less than 
the $\mathcal{O}(N)$ value required for Bose-Einstein condensation.
More generally, $g_{1}(z)$ was found \cite{VaiTra79} to be of order 
$1/\sqrt{k}$ at small $k$. The corresponding momentum distribution is
sharply peaked at low $k$ and falls like $k^{-4}$ at large $k$
\cite{MinVigTos02,OlsDun02,OlsDun03}, very different from the filled Fermi 
sea of the ideal Fermi gas, for which $N(k)$ is unity for $|k|<k_F$ and zero 
for $|k|>k_F$.

The exact ground state of the TG gas is also known in the presence
of a longitudinal trap potential 
$\frac{1}{2}m\omega_{long}^{2}z^2$ \cite{1dsho}.
It follows from the mapping theorem
that the exact N-boson ground state $\psi_{B0}$ is
$\psi_{B0}(z_{1},\cdots,z_{N})=|\psi_{F0}(z_{1},\cdots,z_{N})|$
where $\psi_{F0}$ is the ground state of 
$N$ spinless fermions with the same Hamiltonian and impenetrability constraint.
The fermionic ground state is a Slater determinant of
the lowest $N$ single-particle eigenfunctions $\varphi_n$ of the
harmonic oscillator (HO), where
$\varphi_{n}(z)=\text{const.}e^{-Q^{2}/2}H_{n}(Q)$
with $H_n(Q)$ the Hermite polynomials and
$Q=z/z_{osc}$, $z_{osc}=\sqrt{\hbar/m\omega_{long}}$
being the longitudinal oscillator length.
By factoring the Gaussians out of the determinant and carrying out
elementary row and column operations, one can cancel all terms in each
$H_n$ except the one of highest degree \cite{Aitken}, yielding
a simple but exact analytical expression
of Bijl-Jastrow pair product form for the $N$-boson ground state:
\begin{equation}\label{SHO-ground}
\psi_{B0}(z_{1},\cdots,z_{N})=\text{const.}
\left[\prod_{i=1}^{N}e^{-Q_{i}^{2}/2}\right]
\prod_{1\le j<k\le N}|z_{k}-z_{j}|
\end{equation}
with $Q_{i}=z_{i}/z_{osc}$.
It is interesting to
note the strong similarity between this exact 1D $N$-boson wave
function and the famous Laughlin variational wave function of the 2D ground
state for the quantized fractional Hall effect \cite{Laughlin},
as well as the closely-related wave functions for bosons with weak repulsive
delta-function interactions in a harmonic trap in 2D found 
by Smith and Wilkin \cite{SW}.

Both the single particle density and pair distribution function depend
only on the absolute square of the many-body wave function, and since
$|\psi_{B0}|^{2}=|\psi_{F0}|^{2}$ they reduce to standard ideal Fermi
gas expressions. The single particle density, normalized to $N$, is
$n(z)=\sum_{n=0}^{N-1}|\varphi_{n}(z)|^{2}$ and the 
pair distribution function, normalized to $N(N-1)$, is
\begin{eqnarray}
D(z_{1},z_{2}) & = & n(z_{1})n(z_{2})
-|\Delta(z_{1},z_{2})|^{2} \nonumber\\
\Delta(z_{1},z_{2}) & = &
\sum_{n=0}^{N-1}\varphi_{n}^{*}(z_{1})\varphi_{n}(z_{2})\quad .
\end{eqnarray}
Although the Hermite polynomials have disappeared from
the expression (\ref{SHO-ground}) for the many-body wave function, they 
reappear upon
integrating $|\psi_{B0}|^{2}$ over $(N-1)$ coordinates to get the single
particle density $n(z)$ and over $(N-2)$ to get the pair distribution
function $D(z_{1},z_{2})$, and these expressions in terms of the HO orbitals
$\varphi_n$ are the most convenient ones for evaluation.
Some qualitative features
of the pair distribution function are apparent: In the first place it
vanishes at contact $z_1=z_2$, as it must because of
impenetrability of the particles. Furthermore, the correlation term
$\Delta(z_{1},z_{2})$ is a truncated closure sum and approaches 
$\delta(z_{1}-z_{2})$ as $N\rightarrow\infty$, as is to
be expected since the healing length in a spatially uniform 1D hard core
Bose gas varies inversely with particle number \cite{soliton}.
As a result the width of the null around the diagonal $Q_1=Q_2$
decreases with increasing $N$, and vanishes in the limit. 
For $|z_{1}-z_{2}|$ much larger than
the healing length, $D$ reduces to the uncorrelated density product
$n(z_{1})n(z_{2})$, so the spatial extent of the pair distribution
function is that of the density and varies as $N^{1/2}$ \cite{Kolomeisky}.
Detailed gray-scale plots of $D(z_{1},z_{2})$ in the $(Q_{1},Q_{2})$ plane
for the cases $N=2$, $N=6$, and $N=10$ are shown in Fig. 1 of \cite{1dsho}.

The reduced single-particle density matrix with
normalization $\int\rho_{1}(z,z)dz=N$ is
\begin{eqnarray}
\rho_{1}(z,z')&=&N\int\psi_{B0}(z,z_{2},\cdots,z_{N})\nonumber\\
&&\times\psi_{B0}(z',z_{2},\cdots,z_{N})dz_{2}\cdots dz_{N}\quad .
\end{eqnarray}
For $z\ne z^{'}$ it cannot be expressed in terms of $|\psi_{B0}|^2$,
and is therefore very different from that of the ideal Fermi gas. 
The multi-dimensional integral cannot be evaluated
analytically, but in \cite{1dsho} it was evaluated  numerically by Monte Carlo 
integration for not too large values of $N$, and grayscale plots are shown
in Fig. 2 of \cite{1dsho}. More accurate numerical results
were found in \cite{LapGirWri02}, and highly accurate results for large values
of $N$ were found in \cite{ForFraGarWit03}.
In a macroscopic system, the presence or absence of BEC is determined by the
behavior of $\rho_{1}(z,z')$ as $|z-z'|\rightarrow\infty$. Off-diagonal
long-range order is present if the largest eigenvalue of
$\rho_1$ is macroscopic (proportional to $N$), in which case the system
exhibits BEC and the corresponding eigenfunction, the condensate
orbital, plays the role of an order parameter \cite{PO,Yang2}. Although
this criterion is not strictly applicable to mesoscopic systems, if the
largest eigenvalue of $\rho_1$ is much larger than one
then it is reasonable to expect that
the system will exhibit some BEC-like coherence effects. Thus we examine here
the spectrum of eigenvalues $\lambda_j$ and associated eigenfunctions
$\phi_{j}(z)$ (``natural orbitals'') of $\rho_1$. 
The relevant eigensystem equation is
\begin{equation}
\int_{-\infty}^{\infty}\rho_{1}(z,z')\phi_{j}(z')dz'=\lambda_{j}\phi_{j}(z)
\quad .
\end{equation}
$\lambda_j$ represents the occupation of the orbital $\phi_j$, and
one has $\sum_{j}\lambda_{j}=N$. Accurate values of the $\lambda_j$ have
been determined in \cite{ForFraGarWit03}. In particular, 
the largest eigenvalue $\lambda_1$ was
shown to be of order $\sqrt{N}$ for large $N$, as in the spatially uniform
case. 

\begin{figure}
\includegraphics[width=1.0\columnwidth,angle=0]{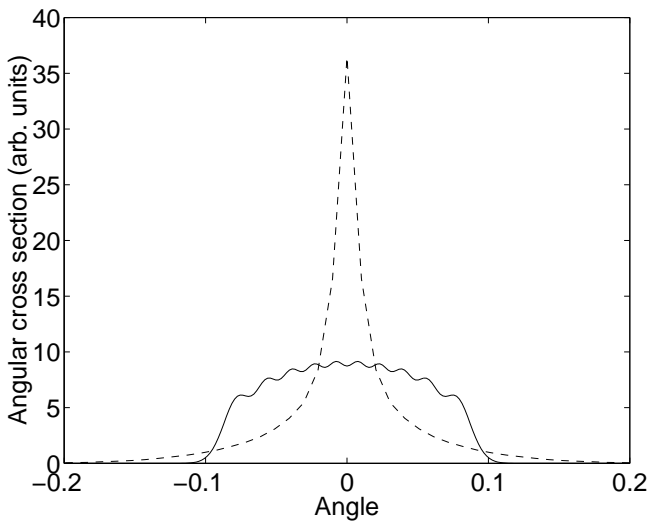}
\caption{Angular
cross section versus angle $\sin(\theta)\approx\theta$ for $N=10$.
The dashed line is for the 1D gas of impenetrable bosons and the
solid line is for the corresponding system of non-interacting
fermions}
\label{Fig:two}
\vspace{-0.5cm}
\end{figure}

Next we examine the momentum distribution, which can be shown \cite{1dsho}
to be a double Fourier transform of $\rho_1$:
\begin{equation}
N(k)=(2\pi)^{-1}\int_{-\infty}^{\infty}dz\int_{-\infty}^{\infty}dz'
\rho_{1}(z,z')e^{-ik(z-z')}\quad .
\end{equation}
The spectral representation of the density matrix then leads to
$N(k)=\sum_{j}\lambda_{j}|\mu_{j}(k)|^2$ where the $\mu_j$ are
Fourier transforms of the natural orbitals:
$\mu_{j}(k)=(2\pi)^{-1/2}\int_{-\infty}^{\infty}\phi_{n}(z)e^{-ikz}dz$.
The key features are that the momentum spectrum maintains the
sharp peaked structure reminiscent of the spatially uniform case
\cite{Ols98,Len64,Len66} and that the peak becomes
sharper with increasing atom number $N$. 
By way of contrast, for a 1D Fermi gas the corresponding momentum
spectrum is a filled Fermi sea and can be expressed as
$N(k)=\sum_{j=1}^N|\mu_{j}(k)|^2$. In
a recent paper \cite{GirWri00} we devised a scheme to measure the momentum
spectrum based on Raman outcoupling and showned that
the angular cross section accurately mirrors the momentum distribution.
Figure \ref{Fig:two} shows 
the angular cross section versus angle for both 
$N=10$ impenetrable bosons 
(dashed line) and the corresponding system of non-interacting
fermions (solid line); see \cite{GirWri00} for details.
\subsection{\label{subsec:FBb}Magnetically trapped, spin-aligned fermions}
Consider first the case $N=2$. The spins are frozen in the configuration
$\uparrow\uparrow$ by the magnetic field, so the spatial relative wave
function $\psi_{F}(z)$ is antisymmetric (odd in $z=z_{1}-z_{2}$). 
The Hamiltonian $\hat{H}_{1D}^{F}$ and corresponding odd-wave pseudopotential
were derived in Sec. \ref{subsubsec:spin-aligned}.
Defining a mapped bosonic (even in $z$) wave function by 
$\psi_{B}(z)=\text{sgn}(z)\psi_{F}(z)$ and mapped scattering length
$a_{1D}^{B}=a_{1D}^{F}\equiv a_{1D}$ 
where $\text{sgn}(z)$ is $+1$ if $z>0$ and $-1$ if $z<0$, one finds that
$\psi_B$ satisfies the usual bosonic contact condition 
$\psi_{B}^{'}(0+)=-\psi_{B}^{'}(0-)=-(a_{1D}^{B})^{-1}\psi_{B}(0\pm)$, 
the zero-range limit $z_{0}\to 0+$ of Eq. (\ref{eq1}).
Since the kinetic
energy contributions from $z\ne 0$ also agree, one has a mapping from
the fermionic to bosonic problem which preserves energy 
eigenvalues and dynamics. The relationship between coupling constants
$g_{1D}^{F}$ in $\hat{H}_{1D}^{F}$ and $g_{1D}^{B}$ in
$\hat{H}_{1D}^{B}=-(\hbar^{2}/2\mu)\partial_{z}^{2}+g_{1D}^{B}\delta(z)$
is $g_{1D}^{B}=\hbar^{4}/\mu^{2}g_{1D}^{F}$, and by
(\ref{Fermi-renorm}) this relationship agrees with the low-energy
limit of Eq. (25) of \cite{GraBlu03,Note1}. 
In the limit $g_{1D}^{B}=+\infty$ arising when $V_{p}\to 0-$, this is
the $N=2$ case of the original mapping \cite{Gir60,Gir65} from hard sphere 
bosons to an ideal Fermi gas, but now generalized to 
arbitrary coupling constants and used in the inverse direction.
This generalizes to arbitrary $N$: Antisymmetric 
fermionic solutions $\psi_{F}(z_{1},\cdots,z_{N})$
are mapped to symmetric bosonic solutions $\psi_{B}(z_{1},\cdots,z_{N})$ via
Eqs. (\ref{eq3}) and (\ref{eq4}). 
The Fermi contact conditions are 
$\psi_{F}|_{z_{j}=z_{\ell}+}=-\psi_{F}|_{z_{j}=z_{\ell}-}
=-(a_{1D}/2)(\partial_{z_{j}}-\partial_{z_{\ell}})
\psi_{F}|_{z_{j}=z_{\ell\pm}}$ and imply the Bose contact conditions
$(\partial_{z_{j}}-\partial_{z_{\ell}})\psi_{B}|_{z_{j}=z_{\ell}+}
=-(\partial_{z_{j}}-\partial_{z_{\ell}})\psi_{B}|_{z_{j}=z_{\ell}-}
=-(2/a_{1D})\psi_{B}|_{z_{j}=z_{\ell}}$, and these are the
usual LL contact conditions \cite{LieLin63}.
The fermionic Hamiltonian is 
$\hat{H}_{1D}^F=-(\hbar^{2}/2\mu)\sum_{j=1}^{N}\partial_{z_j}^{2}
+g_{1D}^{F}\sum_{1\le j<\ell\le N}\delta^{'}(z_{j\ell})
\hat{\partial}_{j\ell}$ where 
$\hat{\partial}_{j\ell}\psi=(1/2)[\partial_{z_{j}}\psi|_{z_{j}=z_{\ell +}}
-\partial_{z_{\ell}}\psi|_{z_{j}=z_{\ell -}}]$.
Although well-defined in the exact Schr\"{o}dinger equation and in
first-order perturbation theory, this fermionic pseudopotential becomes
ambiguous in higher-order perturbation theory. However, after mapping
to the bosonic Hilbert space one has the usual
Lieb-Liniger interaction $g_{1D}^{B}\delta(z_{j\ell})$ which is well-behaved
in all perturbation orders and in second quantization.
This generalization of the Fermi-Bose mapping theorem, due to Cheon and 
Shigehara \cite{CheShi98}, extends the useful domain
of the mapping of Eqs. (\ref{eq3}) and (\ref{eq4}) to the whole range of
coupling constants $g_{1D}^B$ and $g_{1D}^F$. 
The first application to the spin-aligned Fermi gas is due to
Blume and Granger, who were led to the mapping by consideration
of the zero-range limit of a K-matrix formulation \cite{GraBlu03}.
They treated only the case $N=2$ but did
not restrict themselves to the low-energy limit considered here.

\begin{figure}
\includegraphics[width=1.0\columnwidth,angle=0]{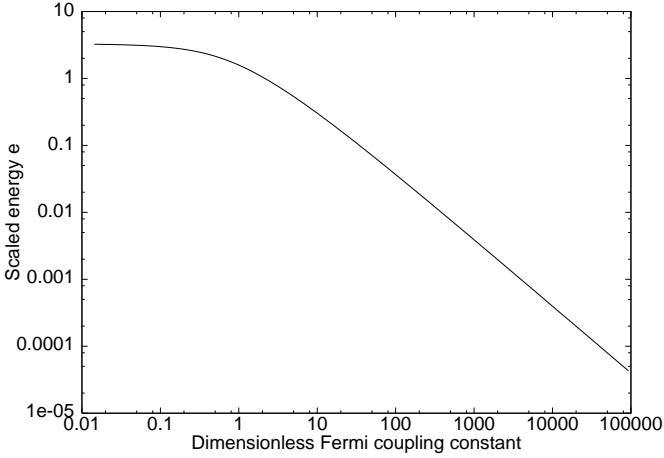}
\caption{Log-log plot of scaled ground state energy per particle 
$e=2m\epsilon/\hbar^{2}n^2$ for the spin-aligned Fermi gas, 
versus dimensionless fermionic coupling constant $\gamma_F$.}
\label{Fig:three}
\vspace{-0.5cm}
\end{figure}

The exact ground state 
\cite{LieLin63} of $\hat{H}_{1D}^B$ is known for all positive
$g_{1D}^B$ if no external potential or nonzero range interactions
are present, and the mapping then generates the exact $N$-body ground
state of $\hat{H}_{1D}^F$. The 
dimensionless bosonic and fermionic coupling constants 
$\gamma_{B}=mg_{1D}^{B}/n\hbar^2$ and $\gamma_{F}=mg_{1D}^{F}n/\hbar^2$  
introduced in Sec. \ref{subsubsec:spin-aligned} satisfy 
$\gamma_{B}\gamma_{F}=4$. 
The energy per particle $\epsilon$ is
related to a dimensionless function $e(\gamma)$ available online
\cite{Note2} via $\epsilon=(\hbar^{2}/2m)n^{2}e(\gamma)$ where $\gamma$
is related to $\gamma_F$ herein by $\gamma=\gamma_{B}=4/\gamma_{F}$.
This is plotted as a function of $\gamma_F$ in Fig. \ref{Fig:three}.
For $g_{1D,F}\to\infty$ as occurs at a p-wave Feshbach resonance,
one has the ``fermionic TG gas'' discussed in Sec. \ref{subsubsec:spin-aligned}
and \cite{GirOls03}. For bosons the TG regime, which maps to the 
\emph{ideal Fermi} gas, is reached when
$g_{1D}^{B}$ is large enough and/or the density $n$ \emph{low} enough that 
$\gamma_{B}\gg 1$. A similar simplification occurs in the fermionic
case, where a fermionic TG regime is reached when $g_{1D}^{F}$ is 
large enough and/or $n$ \emph{high} enough that $\gamma_{F}\gg 1$.
The corresponding fermionic TG gas then maps to the \emph{ideal Bose} gas
since $\gamma_{B}\gamma_{F}=4$.
\subsection{\label{subsec:FBc}Optically trapped, spin-free fermions}
In this section we assume that the Hamiltonian is spin-independent and that
the fermionic vapor is trapped in a tight \emph{optical} atom waveguide. 
The spins are then free to assume whatever configuration minimizes the
ground-state energy. First assume $N=2$. The two-body relative wave functions
are $\psi_{F}(z)=\psi_{F}^{e}(z)+\psi_{F}^{o}(z)$ where 
the spatially even part $\psi_{e}(z)$ contains
an implicit spin-odd singlet spin factor, and the spatially odd part
$\psi_{o}(z)$ contains implicit spin-even triplet spin factors. States of 
combined space-spin bosonic symmetry can be defined by the same mapping 
$\psi_{B}(z)=\text{sgn}(z)\psi_{F}(z)$ used in the previous section,
which now maps the spatially even fermionic function
$\psi_{F}^{e}$ to a spatially odd bosonic function $\psi_{B}^{o}$ and
the spatially odd fermionic function $\psi_{F}^{o}$ to a spatially even
bosonic function $\psi_{B}^{e}$ while leaving the spin dependence unchanged.
Then the even-wave contact conditions for $a_{1D,B}^{e}$ follow from the 
odd-wave contact conditions for $a_{1D,F}^{o}$ and the odd-wave contact 
conditions for $a_{1D,B}^{o}$ follow from the even-wave contact conditions for
$a_{1D,F}^{e}$. As before, one has a mapping from the fermionic to bosonic 
problem which preserves energy eigenvalues and dynamics. The bosonic 
Hamiltonian is of
the same  form as the fermionic one (\ref{H_1D}) but with mapped coupling
constants $g_{1D,B}^{e}=\hbar^{4}/\mu^{2}g_{1D,F}^{o}$ and 
$g_{1D,B}^{o}=\hbar^{4}/\mu^{2}g_{1D,F}^{e}$. This generalizes to
arbitrary $N$: Fermionic solutions 
$\psi_{F}(z_{1},\sigma_{1};\cdots;z_{N},\sigma_{N})$ are mapped to bosonic 
solutions $\psi_{B}(z_{1},\sigma_{1};\cdots;z_{N},\sigma_{N})$ by the
usual mapping (\ref{eq3}), (\ref{eq4}), where the spin z-component variables
$\sigma_j$ take on the values $\uparrow$ and $\downarrow$.
The N-fermion and N-boson Hamiltonians are both of the form
$\hat{H}_{1D}=-(\hbar^2/2m)\sum_{j=1}^{N}\partial_{z_j}^{2}
+\sum_{1\le j<\ell\le N}[g_{1D}^{e}\hat{\delta}_{j\ell}
+g_{1D}^{o}\delta^{'}(z_{j\ell})\hat{\partial}_{j\ell}]$.
On fermionic states $\psi_F$, $g_{1D}^{e}$ and 
$g_{1D}^{o}$ are $g_{1D,F}^{e}$ and $g_{1D,F}^{o}$, whereas on the mapped
bosonic states $\psi_{B}=A\psi_F$ they are 
$g_{1D,B}^{e}=\hbar^{4}/\mu^{2}g_{1D,F}^{o}$ 
and $g_{1D,B}^{o}=\hbar^{4}/\mu^{2}g_{1D,F}^{e}$. This is discussed in
more detail in \cite{GirOls04}.

Assume that both $g_{1D,F}^{e}\ge 0$ and 
$g_{1D,F}^{o}\ge 0$. If $g_{1D,F}^{o}$ is zero  
then it follows from a theorem of Lieb and Mattis \cite{LieMat62} that
the fermionic ground state has total spin $S=0$ (assuming $N$ 
even), as shown in the spatially uniform case by Yang \cite{Yan67} and 
with longitudinal trapping by Astrakharchik {\it et al.} 
\cite{AstBluGioPit03}. 
If $g_{1D,F}^{o}$ is not negligible then the ground state may not have $S=0$. 
In fact, if $g_{1D,F}^{e}$ is zero then one can apply a
theorem of Eisenberg and Lieb \cite{EisLie02} to the mapped spinor boson
Hamiltonian, with the conclusion that one of the degenerate Bose ground 
states is
totally spin-polarized, has $S=N/2$, and is the product of a symmetric
spatial wave function $\psi_{B0}$ and a symmetric spin wave function. 
The ground state is then the same as the one discussed in the previous 
section, except that now there is an $(N+1)$-fold directional degeneracy 
since $S_z$ can range from $-N/2$ to $N/2$. Any $S=0$ state 
has a higher energy; in fact, for $N>2$ the mapped Bose gas is 
partially space-antisymmetric, raising its energy by the exclusion principle.

So far we have considered only the extreme cases where either the even-wave
or odd-wave coupling constant vanishes. Assume now that they may take on
any non-negative values. Consider first the case $N=2$
of a longitudinally trapped spinor Fermi gas, with relative spatial wave
function $\psi_{F}(z)$ and Hamiltonian differing from Eq. (\ref{H_1D}) 
by addition of a harmonic trap potential 
$\frac{1}{2}\mu\omega_{long}^{2}z^2$. The even and odd-wave coupling
constants are $g_{1D,F}^{e} = -\hbar^{2}/\mu a_{1D,F}^e$ and 
$g_{1D,F}^{o} = -\hbar^{2}a_{1D,F}^{o}/\mu$ as before. $\psi_F$ may be 
taken to be either spatially even with associated singlet spin function 
$\frac{1}{\sqrt{2}}(\uparrow\downarrow-\downarrow\uparrow)$ which has $S=0$,
or else spatially odd with associated spin function which is one of the 
$S=1$ triplets $\uparrow\uparrow$, $\downarrow\downarrow$, or 
$\frac{1}{\sqrt{2}}(\uparrow\downarrow+\downarrow\uparrow)$. The singlet 
case has spatially even wave functions identical
with those of trapped $N=2$ bosons. The odd-wave pseudopotential then
projects to zero so the $S=0$ eigenstates are independent of $g_{1D,F}^o$,
and the even-wave pseudopotential reduces to $g_{1D,F}^{e}\delta(z)$.
The exact eigenstates are known
\cite{Cir01}, being of the form $D_{\nu}(|\xi|)$ where $D_\nu$ is a Weber 
(parabolic cylinder) function \cite{Morse,Gradshteyn} and 
$z=\xi\sqrt{\hbar/2\mu\omega_{long}}$. The absolute value in the argument
leads to a cusp at $z=0$ and the LL cusp condition of 
Eq. (\ref{Bose-contact}) and 
\cite{LieLin63} (with $g_{1D,B}$ replaced by $g_{1D,F}^{e}$)
leads to a transcendental equation for the allowed values of $\nu$:
$\Gamma(\frac{1}{2}-\frac{1}{2}\nu)/\Gamma(-\frac{1}{2}\nu)=-\lambda$
in terms of the dimensionless parameter 
$\lambda=g_{1D,F}^{e}/2\hbar\sqrt{\mu/\hbar\omega_{long}}$. The energy
eigenvalues are $E(\nu)=(\nu +\frac{1}{2})\hbar\omega_{long}$, 
the ground state is
that solution for which $\nu$ vanishes as $\lambda\to 0$, and its
energy $E_0$ is a monotonically increasing function of $g_{1D}^{e}$.
Next consider the $S=1$ (triplet) solutions, for which $\psi_F$ is
spatially odd. Then the even-wave pseudopotential projects to zero,
and on carrying out the Fermi-Bose mapping 
$\psi_{B}(z)=\text{sgn}(z)\psi_{F}(z)$ the odd-wave pseudopotential is
changed to $g_{1D,B}^{e}\ \delta(z)$ with 
$g_{1D,B}^{e}=\hbar^{4}/\mu^{2}g_{1D,F}^{o}$. Thus the $S=1$
ground state energy is the same function of $g_{1D,B}^{e}$ that the
$S=0$ ground state energy is of $g_{1D,F}^{e}$, and is therefore a
monotonically \emph{decreasing} function of $g_{1D,F}^{o}$. It follows
that the $S=0$ and $S=1$ ground-state energies are equal on the
hyperbola $g_{1D,F}^{e}g_{1D,F}^{o}=\hbar^{4}/\mu^2$ in the
($g_{1D,F}^{e}\ ,\ g_{1D,F}^{o}$) plane, which forms a phase boundary between
the region where the absolute ground state has $S=0$ and that where it has
$S=1$. For $g_{1D,F}^{e}g_{1D,F}^{o}<\hbar^{4}/\mu^2$ (below the phase
boundary) the ground state has $S=0$ and its energy is independent
of $g_{1D,F}^{o}$, and for $g_{1D,F}^{e}g_{1D,F}^{o}>\hbar^{4}/\mu^2$
(above the phase boundary) it has $S=1$ and its energy is independent
of $g_{1D,F}^{e}$. 

Since it depends only on a symmetry argument and is independent of 
$\omega_{long}$, it is reasonable to conjecture that the above phase 
boundary is valid for all $N$ and for both spatially uniform and
longitudinally trapped spinor Fermi gases. In other words we make the
following

{\it Conjecture}. The ground state of a spinor one-dimensional Fermi gas
undergoes a quantum phase transition at $\gamma_{e} \gamma_{o} = 4$, being
paramagnetic ($S=0$) for
$\gamma_{e} \gamma_{o} < 4$ and ferromagetic, with $S = N/2$ for
$\gamma_{e} \gamma_{o} > 4$; here $\gamma_e$ and $\gamma_o$ are
the dimensionless even and odd-wave coupling constants
$\gamma_{e}=mg_{1D,F}^{e}/n\hbar^2$ and $\gamma_{o}=mg_{1D,F}^{o}n/\hbar^2$.

As a first step in motivating this conjecture, 
it is convenient to change from the representation in terms of space-spin
antisymmetric wave functions 
$\psi_{F}(z_{1},\sigma_{1};\cdots;z_{N},\sigma_{N})$ to a
representation in which states with $S_{z}=k-\frac{1}{2}N$
are represented by functions 
$\varphi_{F}(x_{1},\cdots,x_{k};y_{1},\cdots,y_{N-k})\uparrow_{1}\cdots
\uparrow_{k}\downarrow_{1}\cdots\downarrow_{N-k}$, where we assume $N$
even and $1\le k\le N$. $\varphi_F$ is antisymmetric under permutations
of the z-coordinates $(x_{1},\cdots,x_{k})$ of up-spin atoms and also
under permutations of those $(y_{1},\cdots,y_{N-k})$ of down-spin atoms,
but it has no particular symmetry under exchanges $x_{i}\leftrightarrow y_j$
of up-spin with down-spin atoms. Formally treating up and down-spin particles
as different species can be shown \cite{Mes62,Gir65} to be physically
equivalent to the usual representation 
$\psi_{F}(z_{1},\sigma_{1};\cdots;z_{N},\sigma_{N})$. 
The Hamiltonian is similar to that of Eq. (3) of \cite{AstBluGioPit03},
but also includes odd-wave interactions: 
\begin{eqnarray}\label{2-species}
&&\hat{H}_{F}=-\frac{\hbar^2}{2m}
\left(\sum_{i=1}^{k}\partial_{x_{i}}^{2}
+\sum_{j=1}^{N-k}\partial_{y_{j}}^{2}\right)\nonumber\\
&&+\sum_{1\le p<q\le k}\hat{v}_{1D,F}^{o}(x_{p},x_{q})
+\sum_{1\le p<q\le N-k}\hat{v}_{1D,F}^{o}(y_{p},y_{q})\nonumber\\
&&+\sum_{i=1}^{k}\sum_{j=1}^{N-k}[\hat{v}_{1D,F}^{e}(x_{i},y_{j})
+\hat{v}_{1D,F}^{o}(x_{i},y_{j})]
\end{eqnarray}
where the even and odd-wave pseudopotential operators are those of
Eq.(\ref{H_1D}) expressed in terms of relative coordinates $x_{p}-x_{q}$,
etc. Even-wave interactions $\hat{v}_{1D,F}^{e}(x_{p},x_{q})$ and
$\hat{v}_{1D,F}^{e}(y_{p},y_{q})$ between like-spin particles are absent
because $\varphi_F$ is antisymmetric in both 
$(x_{1},\cdots,x_{k})$ and in $(y_{1},\cdots,y_{N-k})$, so the corresponding
even-wave interactions project to zero. However, both even and 
odd-wave interactions between up and down-spin particles are present
because the states $\varphi_F$ are neither symmetric nor antisymmetric
under exchanges $x_{i}\leftrightarrow y_{j}$ between up and down-spin
atoms. 

The like-spin odd-wave interactions $\hat{v}_{1D,F}^{o}(x_{p},x_{q})$ and
$\hat{v}_{1D,F}^{o}(y_{p},y_{q})$ can be transformed into much simpler
even-wave interactions by a generalization of the previous
Fermi-Bose mapping. Define a mapped wave function $\varphi_B$ of a
\emph{two-component Bose gas} by 
$\varphi_{B}(x_{1},\cdots,x_{k};y_{1},\cdots,y_{N-k})=A(x_{1},\cdots,x_{k})
A(y_{1},\cdots,y_{N-k})\varphi_{F}(x_{1},\cdots,x_{k};y_{1},\cdots,y_{N-k})$
where $A$ is the previously-defined ``unit antisymmetric function''
of Eq. (\ref{eq3}), equal to $\pm 1$ everywhere, antisymmetric in its
arguments, and changing sign here only at ``same-species'' collisions
$x_{p}=x_{q}$ and $y_{p}=y_{q}$. Then $\varphi_B$ satisfies a mapped
Schr\"{o}dinger equation $\hat{H}_{B}\varphi_{B}=E\varphi_{B}$ with the
same eigenvalue $E$ and with
\begin{eqnarray}\label{2-component-Bose}
&&\hat{H}_{B}=-\frac{\hbar^2}{2m}
\left(\sum_{i=1}^{k}\partial_{x_{i}}^{2}+\sum_{j=1}^{N-k}\partial_{y_{j}}^{2}\right)\nonumber\\
&&+\sum_{1\le p<q\le k}g_{1D,B}^{e}\delta(x_{p}-x_{q})\nonumber\\
&&+\sum_{1\le p<q\le N-k}g_{1D,B}^{e}\delta(y_{p}-y_{q})\nonumber\\
&&+\sum_{i=1}^{k}\sum_{j=1}^{N-k}[\hat{v}_{1D,F}^{e}(x_{i},y_{j})
+\hat{v}_{1D,F}^{o}(x_{i},y_{j})]\quad .
\end{eqnarray}
Now there are only interactions of LL type \cite{LieLin63} between like-spin 
atoms, and these are straightforward to treat; they have a mapped coupling
constant $g_{1D,B}^{e}=\hbar^{4}/\mu^{2}g_{1D,F}^{o}$. On the other hand,
the mapping does not affect the contact condition for collisions
$x_{i}=y_{j}$ of unlike-species atoms, so these interactions are the
same as in Eq. (\ref{2-species}) and retain the original fermionic coupling 
constants $g_{1D,F}^{e}$ and $g_{1D,F}^{o}$.

Consider now the Gross-Pitaevskii (GP) approximation, the variational
energy with a trial function 
$\varphi_{B}^{GP}=[\prod_{i=1}^{k}u_{\uparrow}(x_{i})]
[\prod_{j=1}^{N-k}u_{\downarrow}(y_{j})]$. In the absence of longitudinal
trapping one has $u_{\uparrow}(x_{i})=u_{\downarrow}(y_{j})=L^{-1/2}$
where $L$ is the length of the periodic box. When acting on these trivial
wave functions the odd-wave interactions $\hat{v}_{1D,F}^{o}(x_{i},y_{j})$ in
(\ref{2-component-Bose}) project to zero and the even-wave interactions
$\hat{v}_{1D,F}^{e}(x_{i},y_{j})$ reduce to simple Lieb-Liniger interactions
$g_{1D,F}^{e}\delta(x_{i}-y_{j})$, so the variational energy per
particle in the thermodynamic limit ($N\to\infty,L\to\infty,N/L\to n$) is 
\begin{equation}\label{GP-energy}
\epsilon_{0}^{GP}=n[\alpha^{2}(g_{1D,B}^{e}-g_{1D,F}^{e})
+\frac{1}{4}(g_{1D,B}^{e}+g_{1D,F}^{e})] 
\end{equation}
where $\alpha=S_{z}/N$ 
with $S_{z}=k-\frac{1}{2}N$. Thus $\epsilon_{0}^{GP}$ increases with $S_z$ if
$g_{1D,B}^{e}>g_{1D,F}^{e}$ and decreases with $S_z$ if 
$g_{1D,B}^{e}<g_{1D,F}^{e}$. It follows that the ground state has $S_{z}=0$
if $g_{1D,B}^{e}>g_{1D,F}^{e}$ or equivalently 
$g_{1D,F}^{e}g_{1D,F}^{o}<\hbar^{4}/\mu^2$, and $S=S_{z}=\frac{N}{2}$ 
(the maximal value)
if $g_{1D,F}^{e}g_{1D,F}^{o}>\hbar^{4}/\mu^2$. This is the same result
obtained previously for the exact $N=2$ ground state in a trap.  
The ground state has $S_{z}=0$ if 
$\gamma_{e}\gamma_{o}<4$ and $S=S_{z}=\frac{N}{2}$ if 
$\gamma_{e}\gamma_{o}>4$, so there
is a hyperbolic phase boundary $\gamma_{e}\gamma_{o}=4$ in the 
($\gamma_{e}\ ,\ \gamma_o$) plane. In the $S=\frac{N}{2}$ 
($\alpha=\frac{1}{2}$) phase 
the ground state energy depends only on $g_{1D,B}^{e}$, hence on $\gamma_o$
but not on $\gamma_e$, a result which we will show holds also for the
\emph{exact} $N$-atom ground state. 

The GP approximation is valid when both $g_{1D,F}^e$ and $g_{1D,B}^e$
are small enough, more precisely when both $\gamma_{e}\ll 1$ and
$\gamma_{o}\gg 1$. Since the product $\gamma_{e}\gamma_{o}$ can be
made to assume any desired value without violating these inequalities,
it is reasonable to suppose that the phase boundary 
$\gamma_{e}\gamma_{o}=4$ holds for the \emph{exact} ground state. We now show
that this is true. To see this, note first that since $\varphi_B$ is
symmetric in $(x_{1},\cdots,x_{k})$ and in $(y_{1},\cdots,y_{N-k})$,
it follows that when acting on $\varphi_B$, $g_{1D,B}^{e}\delta(x_{p}-x_{q})$
is equivalent to $\hat{v}_{1D,B}^{e}(x_{p},x_{q})$ defined in connection with 
Eq. (\ref{even-interaction}) with $g_{1D}^{e}=g_{1D,B}^{e}$, and
similarly $g_{1D,B}^{e}\delta(y_{p}-y_{q})$ is equivalent to
$\hat{v}_{1D,B}^{e}(y_{p},y_{q})$. Furthermore, one may formally add
interactions $\hat{v}_{1D,F}^{o}(x_{p},x_{q})$ and
$\hat{v}_{1D,F}^{o}(y_{p},y_{q})$ since these vanish on $\varphi_B$
because of its symmetry in $(x_{1},\cdots,x_{k})$ and 
$(y_{1},\cdots,y_{N-k})$. Thus in (\ref{2-component-Bose}) we may replace
$g_{1D,B}^{e}\delta(x_{p}-x_{q})$ by 
$[\hat{v}_{1D,B}^{e}(x_{p},x_{q})+\hat{v}_{1D,F}^{o}(x_{p},x_{q})]$ and
$g_{1D,B}^{e}\delta(y_{p}-y_{q})$ by
$[\hat{v}_{1D,B}^{e}(y_{p},y_{q})+\hat{v}_{1D,F}^{o}(y_{p},y_{q})]$
without changing its action on $\varphi_B$. After this has been done,
we note that when $g_{1D,F}^{e}=g_{1D,B}^{e}$, the resultant Hamiltonian
reduces to that of a \emph{one}-component Bose gas with both even and 
odd-wave interactions, with ground state energy $E_{0}(S_{z})$ 
which is independent of $k$, hence independent of $S_z$. 
Hence, on the line $\gamma_{e}\gamma_{o}=4$ ground states of all 
values of $S_z$ from $0$ to $N/2$ are degenerate, and it is easy to show 
that this remains true if spin-independent longitudinal trap potentials are 
added to $\hat{H}_B$. To complete the proof we need to convert this into
a statement about the total spin $S$ of the ground state, not merely its
value of $S_z$. To do this, first map back to the two-component Fermi gas
states 
$\varphi_{F}(x_{1},\cdots,x_{k};y_{1},\cdots,y_{N-k})\uparrow_{1}\cdots
\uparrow_{k}\downarrow_{1}\cdots\downarrow_{N-k}$. They are not in general
eigenstates of $S$, but they can be converted into such eigenstates $\psi_F$,
with the same eigenvalues of both energy and $S_z$, by antisymmetrizing
with respect to all combined space-spin exchanges 
$(z_{i},\sigma_{i})\leftrightarrow(z_{j},\sigma_{j})$ after renaming the 
variables as follows: 
$(x_{1},\cdots,x_{k};y_{1},\cdots,y_{N-k})\to (z_{1},\cdots,z_{N})$ and
$(\uparrow_{1}\cdots\uparrow_{k}\downarrow_{1}\cdots\downarrow_{N-k})
\to (\uparrow_{1}\cdots\uparrow_{k}\downarrow_{k+1}\cdots\downarrow_{N})$. 
In this standard representation the Hamiltonian, total spin, and its
z-component are mutually commuting, so nondegerate energy eigenstates are 
also eigenstates of $\hat{S}$ and $\hat{S}_z$, and degenerate ones can be
chosen to be such simultaneous eigenstates. Let 
$\psi_{F0}(\gamma_{e},\gamma_{o},S_{z},S)$ be the
lowest such state for given values of $\gamma_{e}$ and $\gamma_{o}$.
We have shown that if $\gamma_{e}\gamma_{o}<4$ then such a ground state
has $S_{z}=0$, if $\gamma_{e}\gamma_{o}>4$ it has $S_{z}=\frac{N}{2}$,
and if $\gamma_{e}\gamma_{o}=4$ it can have any value of $S_z$ from
$0$ to $\frac{N}{2}$. If $\gamma_{e}\gamma_{o}>4$ this state also has
$S=\frac{N}{2}$ since $S_{z}=\frac{N}{2}$ implies $S=\frac{N}{2}$.
If $\gamma_{e}\gamma_{o}<4$ then the ground state also has $S=0$ 
because if it had $S>0$  than there would exist states
with $0 < |S_{z}| \le S$ of the same energy, contradicting the
monotonic dependence of energy on $S_{z}$ demonstrated in (\ref{GP-energy}),
which we assume to hold also for the \emph{exact} ground state. 
It follows that the hyperbola
$\gamma_{e}\gamma_{o}=4$ is the boundary between a ground state $S=0$
phase ($\gamma_{e}\gamma_{o}<4$) and a $S=\frac{N}{2}$ phase
($\gamma_{e}\gamma_{o}>4$). 
\section{Prospect and critique}
It follows from the above that the dependence of the ground-state energy
on total spin $S$ is more complicated than envisioned in
either \cite{AstBluGioPit03} or \cite{GirOls04}: It is indeed true that
if $\gamma_{o}$ is \emph{exactly} zero, then the ground state has
$S=0$ as assumed in \cite{AstBluGioPit03}, but if $\gamma_{o}\ne 0$
then by approaching the confinement-induced resonance (CIR) of
$\gamma_e$ implied by Eq. (\ref{CIR-Bose}), one can in principle always make
$\gamma_{e}\gamma_{o}>4$ even if $0<\gamma_{o}\ll 1$, thereby inducing
a phase transition from the paramagnetic $S=0$ phase to the ferromagnetic
$S=N/2$ phase. In the opposite case $\gamma_{o}\gg 1$ encountered near
the CIR of $\gamma_o$ implied by Eq. (\ref{Fermi-renorm}) a phase transition
in the opposite direction ($S=N/2$ to $S=0$) would in principle occur 
if $\gamma_{e}\to 0+$, but in practice there is no way of making
$\gamma_e$ that small. In the region $\gamma_{e}\gamma_{o}>4$ where $S=N/2$ 
the exact ground state is totally spatially antisymmetric and hence 
spin-aligned, so its energy is independent of $\gamma_e$ and given by the 
results in \cite{GirOls04} and Sec. \ref{subsubsec:spin-aligned} herein.
In the region $\gamma_{e}\gamma_{o}<4$ where $S=0$ the ground state is 
thus far only known for the case $\gamma_{o}=0$, where it was determined by 
Yang \cite{Yan67} in the spatially uniform case and by Astrakharchik 
{\it et al.} \cite{AstBluGioPit03} in the longitudinally trapped case. 
No analytical or numerical results are yet known for the ground-state energy 
in the region $\gamma_{e}\gamma_{o}<4$ if $\gamma_o\ne 0$, but it should 
be investigated by numerical calculations. These will be more
complicated than the previous ones \cite{AstBluGioPit03} due to the presence
of both even and odd-wave interactions between up and down-spin atoms.
However, if the ground state is real and nodeless \cite{Note3} they should
be feasible, perhaps by using an interaction potential consisting of a
narrow and deep well to represent the odd-wave interaction (see 
Sec. \ref{subsubsec:spin-aligned}) together with a somewhat broader 
``soft rod'' potential to represent the even-wave interaction.

Our proof of the exact ground-state phase boundary $\gamma_{e}\gamma_{o}=4$
is a ``physicist's proof'' and we make no claim of mathematical rigor.
In the first place, as pointed out in Sec. \ref{subsubsec:p-wave},
a rigorous derivation of the zero-range, 1D  limit of the odd-wave 
interaction between fermions in tight waveguides does not exist at present, 
although we believe that Eq. (\ref{Fermi-renorm}) is a correct zero-range, 
low-energy consequence of the K-matrix treatment of Granger and Blume 
\cite{GraBlu03}. Furthermore, for $N>2$ our proof of the phase boundary
is not completely rigorous since we had to assume that the ground state
energy is a single-valued function of $S_z$ both for $\gamma_{e}\gamma_{o}<4$
and for $\gamma_{e}\gamma_{o}>4$.
We feel that this is justified since we proved it to be true in the GP
approximation and showed that the criteria for validity of the GP 
approximation can be satisfied on the phase boundary $\gamma_{e}\gamma_{o}=4$.
However, a truly rigorous proof of this phase boundary does not yet exist,
and we hope that one will be forthcoming. 
\begin{acknowledgments}
We are very grateful to Doerte Blume for helpful comments and for 
communications regarding her closely-related
works with Brian Granger \cite{GraBlu03} and with Astrakharchik {\it et al.} 
\cite{AstBluGioPit03}, and to Ewan Wright for references
\cite{Ton36,Bij37,Nag40,Sta60}. We thank Michael Moore and Thomas Bergeman for
their invaluable assistance in preparation of the Les Houches lecture
notes \cite{MooBerOls04} on which Sec. \ref{sec:Atomic} is based. 
This work was supported by Office of Naval Research grant N00014-03-1-0427
(M.D.G. and M.O.) and by NSF grants PHY-0301052 and PHY-0070333 (M.O.).
\end{acknowledgments}

\begin{thebibliography}{81}
\bibitem{Cir01} M.A. Cirone, K. Goral, K. Rzazewski, and M. Wilkens, 
J. Phys. B {\bf 34}, 4571 (2001). Note, however, that their Fig. 3 is 
incorrect, since the curves for $c\ne 0$ do not have the necessary cusps
at $z=0$ ; for example, the $c=\infty$ curve (TG limit) should have
its left half reflected up, yielding an even function with the necessary
cusp.
\end{thebibliography}
\end{document}